\documentclass[pdftex,twocolumn,epjc3_preprint,runningheads]{svjour3}

\usepackage[T1]{fontenc}
\usepackage{lmodern}
\usepackage{calc}
\usepackage{graphicx}
\usepackage{booktabs}
\usepackage{textcomp}
\usepackage{xspace}
\usepackage{relsize}
\usepackage{amssymb}
\usepackage{amsmath}
\usepackage{listings}
\usepackage{microtype}
\usepackage{multirow}
\usepackage{tabularx}
\usepackage{array}
\usepackage{placeins}
\usepackage{cuted}
\usepackage{soul} % only for \st; delete if this causes you problems.
\usepackage{fixltx2e}
\usepackage{slashed}
\usepackage{bm}
\usepackage[numbers,sort&compress]{natbib}
\usepackage[labelfont=bf,font=small]{caption}
\usepackage[skip=-2pt]{subcaption}
\usepackage[colorlinks,citecolor=blue,urlcolor=blue,linkcolor=blue,breaklinks=breakall]{hyperref}
\usepackage{breakurl}
\usepackage[dvipsnames]{xcolor}
\usepackage[clockwise,figuresright]{rotating}
\usepackage{siunitx}
\usepackage{tikz}
\usepackage[normalem]{ulem}
\usepackage[utf8]{inputenc}

\usepackage{etoolbox}
\AfterEndEnvironment{strip}{\leavevmode}

\allowdisplaybreaks

\newcolumntype{L}{>{\raggedright\let\newline\\\arraybackslash\hspace{0pt}}X}
\newcolumntype{R}{>{\raggedleft\let\newline\\\arraybackslash\hspace{0pt}}X}
\newcolumntype{C}{>{\centering\let\newline\\\arraybackslash\hspace{0pt}}X}

\newcommand{\gambitinstitute}[1]{\expandafter\csname #1\endcsname \label{#1}}
\newcommand{\gi}[1]{\gambitinstitute{#1} \and }
\newcommand{\last}[1]{\gambitinstitute{#1}}

\makeatletter

\newcommand{\preprintnumber}[1]{\gdef\@preprintnumber{\begin{flushright}{#1}\end{flushright}}}

\g@addto@macro\bfseries{\boldmath}
\makeatother

\newcommand{\subparagraph}{} %< workaround for svjour not defining subparagraph
\usepackage{titlesec}
\titleformat*{\paragraph}{\bfseries}
\journalname{Eur. Phys. J. C}
\smartqed
\let\underscore\_
\renewcommand{\_}{\discretionary{\underscore}{}{\underscore}}

\makeatletter
\let\orgdescriptionlabel\descriptionlabel
\renewcommand*{\descriptionlabel}[1]{%
  \let\orglabel\label
  \let\label\@gobble
  \phantomsection
  \protected@edef\@currentlabel{#1}%
  \let\label\orglabel
  \orgdescriptionlabel{#1}%
}
\makeatother

\newcommand\postnewlinemarker{\hbox{\ensuremath{\hookrightarrow}}}
\newcommand\cpp[1]{{\lstinline!#1!}}  % Apparently curly braces are only "experimental"

\newcommand\yaml[1]{{\lstset{style=yaml}\lstinline!#1!\lstset{style=cpp}}}

\newcommand\term[1]{{\lstset{style=terminal}\lstinline!#1!\lstset{style=cpp}}}
\newcommand\fortran[1]{{\lstset{style=fortran}\lstinline!#1!\lstset{style=cpp}}}
\newcommand\py[1]{{\lstset{style=python}\lstinline!#1!\lstset{style=cpp}}}
\newcommand\customtilde{{\raisebox{0.2ex}{\scalebox{0.6}{\boldmath$\sim$}}}}
\newcommand\mathematica[1]{{\lstset{style=Mathematica}\lstinline!#1!\lstset{style=cpp}}}

\lstnewenvironment{lstlistingyaml}{\lstset{style=yaml}}{\lstset{style=cpp}}
\lstnewenvironment{lstlistingterm}{\lstset{style=terminal}}{\lstset{style=cpp}}
\lstnewenvironment{lstlistingfortran}{\lstset{style=fortran}}{\lstset{style=cpp}}
\lstnewenvironment{lstcpp}{\lstset{style=cpp}}{\lstset{style=cpp}}
\lstnewenvironment{lstcppalt}{\lstset{style=cppalt}}{\lstset{style=cpp}}
\lstnewenvironment{lstcppnum}{\lstset{style=cppnum}}{\lstset{style=cpp}}
\lstnewenvironment{lstyaml}{\lstset{style=yaml}}{\lstset{style=cpp}}
\lstnewenvironment{lstterm}{\lstset{style=terminal}}{\lstset{style=cpp}}
\lstnewenvironment{lsttermalt}{\lstset{style=terminalalt}}{\lstset{style=cpp}}
\lstnewenvironment{lsttext}{\lstset{style=text}}{\lstset{style=cpp}}
\lstnewenvironment{lstfortran}{\lstset{style=fortran}}{\lstset{style=cpp}}
\lstnewenvironment{lstpy}{\lstset{style=python}}{\lstset{style=cpp}}
\lstnewenvironment{lstmathematica}{\lstset{style=mathematica}}{\lstset{style=cpp}}

\newcommand{\tmpname}{}
\newcommand{\tmplistingname}{}
\makeatletter
\newif\ifATOlabelname
\lst@Key{labelname}{Listing}{\def\ATOlabelname{#1}\global\ATOlabelnametrue}
\makeatother
\lstnewenvironment{lstcpplabel}[1][]{
  \lstset{style=cpp,#1} % #1 allows to add new options with [] same as for normal lstlistings environment
  \ifATOlabelname
    \renewcommand{\tmpname}{\lstlistingname}
    \renewcommand{\tmplistingname}{\lstlistlistingname}
    \renewcommand{\lstlistingname}{\ATOlabelname}% Listing -> labelname
    \renewcommand{\lstlistlistingname}{List of \lstlistingname s}% List of Listings -> List of labelname
  \fi
}{
  \renewcommand{\lstlistingname}{\tmpname}
  \renewcommand{\lstlistlistingname}{\tmplistingname}
  \lstset{style=cpp}
}
\definecolor{solarized@base03}{HTML}{002B36}
\definecolor{solarized@base02}{HTML}{073642}
\definecolor{solarized@base01}{HTML}{586e75}
\definecolor{solarized@base00}{HTML}{657b83}
\definecolor{solarized@base0}{HTML}{839496}
\definecolor{solarized@base1}{HTML}{93a1a1}
\definecolor{solarized@base2}{HTML}{EEE8D5}
\definecolor{solarized@base3}{HTML}{FDF6E3}
\definecolor{solarized@yellow}{HTML}{B58900}
\definecolor{solarized@orange}{HTML}{CB4B16}
\definecolor{solarized@red}{HTML}{DC322F}
\definecolor{solarized@magenta}{HTML}{D33682}
\definecolor{solarized@violet}{HTML}{6C71C4}
\definecolor{solarized@blue}{HTML}{268BD2}
\definecolor{solarized@cyan}{HTML}{2AA198}
\definecolor{solarized@green}{HTML}{859900}
\definecolor{darkred}{HTML}{550003}
\definecolor{darkgreen}{HTML}{00AA00}

\newcommand\YAMLstringstyle{\footnotesize\color{solarized@green}\mdseries}
\newcommand\YAMLkeystyle{\footnotesize\color{solarized@blue}\ttfamily}
\newcommand\YAMLvaluestyle{\footnotesize\color{blue}\mdseries}
\newcommand\ProcessThreeDashes{\llap{\color{cyan}\mdseries-{-}-}}
\newcommand\CPPcommentstyle{\color{solarized@violet}\footnotesize\ttfamily}
\newcommand\CPPdirectivestyle{\color{solarized@magenta}\footnotesize\ttfamily}
\newcommand\termplainstyle{\footnotesize\ttfamily}

\newcommand\processLongMacroDelimiter
{%
\CPPdirectivestyle%
\#define%
}

\lstdefinestyle{cpp}
{
  language=C++,
  basicstyle=\footnotesize\ttfamily,
  basewidth={0.53em,0.44em}, %Ben: experimenting a bit with the fixed-width width (first argument); feels a bit more readable to me with the slightly smaller width (was 0.6em by default)
  numbers=none,
  tabsize=2,
  breaklines=true,
  escapeinside={@}{@},
  showstringspaces=false,
  numberstyle=\tiny\color{solarized@base01},
  keywordstyle=\color{solarized@orange},
  stringstyle=\color{solarized@red}\ttfamily,
  identifierstyle=\color{solarized@blue},
  commentstyle=\CPPcommentstyle,
  directivestyle=\CPPdirectivestyle,
  emphstyle=\color{solarized@green},
  frame=single,
  rulecolor=\color{solarized@base2},
  rulesepcolor=\color{solarized@base2},
  literate={~} {\customtilde}1,
  moredelim=*[directive]\ \ \#,
  moredelim=*[directive]\ \ \ \ \#
}

\lstdefinestyle{cppalt}
{
  language=C++,
  basicstyle=\footnotesize\ttfamily,
  basewidth={0.53em,0.44em}, %Ben: experimenting a bit with the fixed-width width (first argument); feels a bit more readable to me with the slightly smaller width (was 0.6em by default)
  numbers=none,
  tabsize=2,
  breaklines=true,
  escapeinside={*@}{@*},
  showstringspaces=false,
  numberstyle=\tiny\color{solarized@base01},
  keywordstyle=\color{solarized@orange},
  stringstyle=\color{solarized@red}\ttfamily,
  identifierstyle=\color{solarized@blue},
  commentstyle=\CPPcommentstyle,
  directivestyle=\CPPdirectivestyle,
  emphstyle=\color{solarized@green},
  frame=single,
  rulecolor=\color{solarized@base2},
  rulesepcolor=\color{solarized@base2},
  literate={~}{\customtilde}1,
  moredelim=**[is][\processLongMacroDelimiter]{BeginLongMacro}{EndLongMacro} %special delimiter for long macros that go over several lines
}

\lstdefinestyle{cppnum}
{
  language=C++,
  basicstyle=\footnotesize\ttfamily,
  basewidth={0.53em,0.44em}, %Ben: experimenting a bit with the fixed-width width (first argument); feels a bit more readable to me with the slightly smaller width (was 0.6em by default)
  numbers=none,
  tabsize=2,
  breaklines=true,
  escapeinside={@}{@},
  numberstyle=\tiny\color{solarized@base01},
  showstringspaces=false,
  numberstyle=\tiny\color{solarized@base01},
  keywordstyle=\color{solarized@orange},
  stringstyle=\color{solarized@red}\ttfamily,
  identifierstyle=\color{solarized@blue},
  commentstyle=\CPPcommentstyle,
  directivestyle=\CPPdirectivestyle,
  emphstyle=\color{solarized@green},
  frame=single,
  rulecolor=\color{solarized@base2},
  rulesepcolor=\color{solarized@base2},
  literate={~} {\customtilde}1,
  moredelim=*[directive]\ \ \#,
  moredelim=*[directive]\ \ \ \ \#
}

\lstdefinestyle{python}
{
  language=Python,
  basicstyle=\footnotesize\ttfamily,
  basewidth={0.53em,0.44em},
  numbers=none,
  tabsize=2,
  breaklines=true,
  escapeinside={@}{@},
  showstringspaces=false,
  numberstyle=\tiny\color{solarized@base01},
  keywordstyle=\color{blue},
  stringstyle=\color{orange}\ttfamily,
  identifierstyle=\color{darkred},
  commentstyle=\color{purple},
  emphstyle=\color{green},
  frame=single,
  rulecolor=\color{solarized@base2},
  rulesepcolor=\color{solarized@base2},
  literate = {~}{\customtilde}1
             {\ as\ }{{\color{blue}\ as\ \color{black}}}3
}

\lstdefinestyle{fortran}
{
  language=Fortran,
  basicstyle=\footnotesize\ttfamily,
  basewidth={0.53em,0.44em},
  numbers=none,
  tabsize=2,
  breaklines=true,
  escapeinside={@}{@},
  showstringspaces=false,
  numberstyle=\tiny\color{solarized@base01},
  keywordstyle=\color{blue},
  stringstyle=\color{orange}\ttfamily,
  identifierstyle=\color{Periwinkle},
  commentstyle=\color{purple},
  emphstyle=\color{green},
  morekeywords={and, or, true, false},
  frame=single,
  rulecolor=\color{solarized@base2},
  rulesepcolor=\color{solarized@base2},
  literate={~}{\customtilde}1
}

\lstdefinestyle{terminal}
{
  language=bash,
  basicstyle=\termplainstyle,
  numbers=none,
  tabsize=2,
  breaklines=true,
  escapeinside={@}{@},
  frame=single,
  showstringspaces=false,
  numberstyle=\tiny\color{solarized@base01},
  keywordstyle=\color{solarized@orange},
  stringstyle=\color{solarized@red}\ttfamily,
  identifierstyle=\color{black},
  commentstyle=\color{solarized@violet},
  emphstyle=\color{solarized@green},
  frame=single,
  rulecolor=\color{solarized@base2},
  rulesepcolor=\color{solarized@base2},
  morekeywords={gambit, cmake, make, mkdir},
  deletekeywords={test},
  literate = {\ gambit}{{\ }{\color{black}}gambit}7
             {/gambit}{{/}{\color{black}}gambit}6
             {gambit/}{{\color{black}}gambit{/}}6
             {/include}{{/}{\color{black}}include}8
             {cmake/}{{\color{black}}cmake/}6
             {.cmake}{{.}{\color{black}}cmake}6
             {~}{\customtilde}1
}

\lstdefinestyle{terminalalt}
{
  language=bash,
  basicstyle=\footnotesize\ttfamily,
  numbers=none,
  tabsize=2,
  breaklines=true,
  escapeinside={*@}{@*},
  frame=single,
  showstringspaces=false,
  numberstyle=\tiny\color{solarized@base01},
  keywordstyle=\color{solarized@orange},
  stringstyle=\color{solarized@red}\ttfamily,
  identifierstyle=\color{black},
  commentstyle=\color{solarized@violet},
  emphstyle=\color{solarized@green},
  frame=single,
  rulecolor=\color{solarized@base2},
  rulesepcolor=\color{solarized@base2},
  morekeywords={gambit, cmake, make, mkdir},
  deletekeywords={test},
  literate = {\ gambit}{{\ }{\color{black}}gambit}7
             {/gambit}{{/}{\color{black}}gambit}6
             {gambit/}{{\color{black}}gambit{/}}6
             {/include}{{/}{\color{black}}include}8
             {cmake/}{{\color{black}}cmake/}6
             {.cmake}{{.}{\color{black}}cmake}6
             {~}{\customtilde}1
}

\lstdefinestyle{text}
{
  language={},
  basicstyle=\footnotesize\ttfamily,
  identifierstyle=\color{black},
  numbers=none,
  tabsize=2,
  breaklines=true,
  escapeinside={*@}{@*},
  showstringspaces=false,
  frame=single,
  rulecolor=\color{solarized@base2},
  rulesepcolor=\color{solarized@base2},
  literate={~}{\customtilde}1
}

\lstdefinestyle{yaml}
{
  language=bash,
  escapeinside={@}{@},
  keywords={true,false,null},
  otherkeywords={},
  keywordstyle=\color{solarized@base0}\bfseries,
  basicstyle=\footnotesize\color{black}\ttfamily,
  identifierstyle=\YAMLkeystyle,
  sensitive=false,
  commentstyle=\color{solarized@orange}\ttfamily,
  morecomment=[l]{\#},
  morecomment=[s]{/*}{*/},
  stringstyle=\YAMLstringstyle\ttfamily,
  moredelim=**[s][\YAMLkeystyle]{,}{:},   % switch to value style at : but back to key style at,
  moredelim=**[l][\YAMLvaluestyle]{:},    % switch to value style at :
  morestring=[b]',
  morestring=[b]",
  literate =    {---}{{\ProcessThreeDashes}}3
                {>}{{\textcolor{solarized@red}\textgreater}}1
                {|}{{\textcolor{solarized@red}\textbar}}1
                {\ -\ }{{\mdseries\color{black}\ -\ \negmedspace}}3
                {\}}{{{\color{black} \}}}}1
                {\{}{{{\color{black} \{}}}1
                {[}{{{\color{black} [}}}1
                {]}{{{\color{black} ]}}}1
                {~}{\customtilde}1,
  breakindent=0pt,
  breakatwhitespace,
  columns=fullflexible
}

\lstdefinestyle{mathematica}
{
  language={Mathematica},
  basicstyle=\footnotesize\ttfamily,
  basewidth={0.53em,0.44em},
  numbers=none,
  tabsize=2,
  breaklines=true,
  escapeinside={@}{@},
  numberstyle=\tiny\color{black},
  showstringspaces=false,
  numberstyle=\tiny\color{solarized@base01},
  keywordstyle=\color{solarized@orange},
  stringstyle=\color{solarized@red}\ttfamily,
  identifierstyle=\color{solarized@orange}\ttfamily,
  commentstyle=\color{solarized@gray}\ttfamily,
  directivestyle=\color{solarized@orange}\ttfamily,
  emphstyle=\color{solarized@green},
  frame=single,
  rulecolor=\color{solarized@base2},
  rulesepcolor=\color{solarized@base2},
  literate={~} {\customtilde}1,
  moredelim=*[directive]\ \ \#,
  moredelim=*[directive]\ \ \ \ \#,
  mathescape=true
}

\newcommand{\doublecross}[2]{\hyperref[#2]{\textbf{#1}}}
\newcommand{\doublecrosssf}[2]{\hyperref[#2]{\textbf{\textsf{#1}}}}

\newcommand{\startglossary}{\section{Glossary}\label{glossary}Here we explain some terms that have specific technical definitions in \GB.\begin{description}}
\newcommand{\finishglossary}{\end{description}}

\newcommand{\bcode}{\begin{lstlisting}}
\newcommand{\ecode}{\end{lstlisting}}

\newcommand{\mh}{m_h}

\newcommand{\MSBar}{\overline{MS}}

\newcommand{\gambit}{\textsf{GAMBIT}\xspace}

\newcommand{\darkbit}{\textsf{DarkBit}\xspace}

\newcommand{\decaybit}{\textsf{DecayBit}\xspace}

\newcommand{\GB}{\gambit}

\newcommand{\ds}{\textsf{DarkSUSY}\xspace}
\newcommand{\darksusy}{\ds}
\newcommand{\wimpsim}{\textsf{WimpSim}\xspace}

\newcommand\nulike{\textsf{nulike}\xspace}
\newcommand\gamLike{\textsf{gamLike}\xspace}
\newcommand\gamlike{\gamLike}
\newcommand{\capgen}{\textsf{Capt'n General}\xspace}

\newcommand\pippi{\textsf{pippi}\xspace}
\newcommand\MultiNest{\textsf{MultiNest}\xspace}
\newcommand\multinest{\MultiNest}

\newcommand\twalk{\textsf{T-Walk}\xspace}
\newcommand\diver{\textsf{Diver}\xspace}
\newcommand\ddcalc{\textsf{DDCalc}\xspace}

\newcommand\beq{\begin{equation}}
\newcommand\eeq{\end{equation}}

\renewcommand{\url}[1]{\href{#1}{#1}}

\pdfoutput=1

\makeatletter
\patchcmd{\ttlh@hang}{\parindent\z@}{\parindent\z@\leavevmode}{}{}
\patchcmd{\ttlh@hang}{\noindent}{}{}{}
\makeatother

\usepackage{mathrsfs}

\newcommand{\ovr}{\overline}

\newcommand{\lagr}{\mathcal{L}}
\newcommand{\lagrsm}{\mathcal{L}_{\textrm{SM}}}
\newcommand{\vo}{v_0}

\newcommand{\ddc}{\textsf{DDCalc 2.0.0}\xspace}

\definecolor{af}{rgb}{0.0, 0.45, 1.0}

\begin{document}

\preprintnumber{ADP-18-22/T1070, COEPP-MN-18-6, DESY-18-141, TTK-18-34}

\title{Global analyses of Higgs portal singlet dark matter models using GAMBIT}

\author
{
The GAMBIT Collaboration:
Peter Athron\thanksref{monash,coepp} \and
Csaba Bal\'azs\thanksref{monash,coepp} \and
Ankit Beniwal\thanksref{coepp,adelaide,okc,su,e1} \and
Sanjay Bloor\thanksref{imperial,e2} \and
Jos\'e Eliel Camargo-Molina\thanksref{imperial} \and
Jonathan M.~Cornell\thanksref{mcgill} \and
Ben Farmer\thanksref{imperial} \and
Andrew Fowlie\thanksref{monash,coepp,nanjing} \and
Tom\'as E.\ Gonzalo\thanksref{oslo} \and
Felix Kahlhoefer\thanksref{aachen,e3} \and
Anders Kvellestad\thanksref{imperial,oslo} \and
Gregory D.\ Martinez\thanksref{ucla} \and
Pat Scott\thanksref{imperial} \and
Aaron C. Vincent\thanksref{queens} \and
Sebastian Wild\thanksref{desy,e4} \and
Martin White\thanksref{coepp,adelaide} \and
Anthony G. Williams\thanksref{coepp,adelaide}
}

\institute{%
  \gi{monash}
  \gi{coepp}
  \gi{adelaide}
  \gi{okc}
  \gi{su}
  \gi{imperial}
  \gi{mcgill}
  \gi{nanjing}
  \gi{oslo}
  \gi{aachen}
  \gi{ucla}
  \gi{queens}
  \last{desy}
}

\thankstext{e1}{ankit.beniwal@fysik.su.se, \href{https://orcid.org/0000-0003-4849-0611}{0000-0003-4849-0611}}
\thankstext{e2}{sanjay.bloor12@imperial.ac.uk}
\thankstext{e3}{kahlhoefer@physik.rwth-aachen.de}
\thankstext{e4}{sebastian.wild@desy.de}

\titlerunning{Global analyses of Higgs portal singlet dark matter models}
\authorrunning{The GAMBIT Collaboration}

\date{Received: date / Accepted: date}

\maketitle

\begin{abstract}
We present global analyses of effective Higgs portal dark matter models in the frequentist and Bayesian statistical frameworks. Complementing earlier studies of the scalar Higgs portal, we use \gambit to determine the preferred mass and coupling ranges for models with vector, Majorana and Dirac fermion dark matter. We also assess the relative plausibility of all four models using Bayesian model comparison. Our analysis includes up-to-date likelihood functions for the dark matter relic density, invisible Higgs decays, and direct and indirect searches for weakly-interacting dark matter including the latest XENON1T data. We also account for important uncertainties arising from the local density and velocity distribution of dark matter, nuclear matrix elements relevant to direct detection, and Standard Model masses and couplings. In all Higgs portal models, we find parameter regions that can explain all of dark matter and give a good fit to all data. The case of vector dark matter requires the most tuning and is therefore slightly disfavoured from a Bayesian point of view. In the case of fermionic dark matter, we find a strong preference for including a CP-violating phase that allows suppression of constraints from direct detection experiments, with odds in favour of CP violation of the order of 100:1. Finally, we present \ddcalc \textsf{2.0.0}, a tool for calculating direct detection observables and likelihoods for arbitrary non-relativistic effective operators.
\end{abstract}

\tableofcontents

%%%%%%%%%%%%%%%%%%%%%%%%%%%%%%%%%%%%%%%%%%%%%%%%%%%%%%%%%%%%%%%%%%%%%%%%%%%%%%%%%%%%%%%%%%%%%%%%%%%%%%%%%%%%%%%%%%%%%%%%
\section{Introduction}

Cosmological and astrophysical experiments have provided firm evidence for the existence of dark matter (DM) \cite{Zwicky33,Rubin70,bullet,Ade:2015xua}.  While the nature of the DM particles and their interactions remains an open question, it is clear that the viable candidates must lie in theories beyond the Standard Model (BSM). A particularly interesting class of candidates are weakly interacting massive particles (WIMPs) \cite{Bertone:2004pz}. They appear naturally in many BSM theories, such as supersymmetry (SUSY) \cite{Jungman:1995df}. Due to their weak-scale interaction cross-section, they can accurately reproduce the observed DM abundance in the Universe today.

So far there is no evidence that DM interacts with ordinary matter in any way except via gravity. However, the generic possibility exists that Standard Model (SM) particles may connect to new particles via the lowest-dimension gauge-invariant operator of the SM, $H^\dagger H$.  It is therefore natural to assume that the standard Higgs boson (or another scalar that mixes with the Higgs) couples to massive DM particles via such a `Higgs portal' \cite{Silveira:1985rk, McDonald:1993ex, Burgess:2000yq, Davoudiasl:2004be, Patt:2006fw, Kim:2008pp, Andreas:2010dz, Aoki:2011zz, Kanemura:2011nm, Raidal:2011xk, Mambrini:2011ik, He:2011de, Drozd:2011aa, Djouadi:2011aa, Nabeshima:2012pz, Okada:2012cc, LopezHonorez:2012kv, Djouadi:2012zc, Walker:2013hka, Okada:2013bna, Chacko:2013lna}. The discovery of the Higgs boson in 2012 by ATLAS \cite{Aad:2012tfa} and CMS \cite{Chatrchyan:2012xdj} therefore opens an exciting potential window for probing DM.

Despite being simple extensions of the SM in terms of particle content and interactions, Higgs portal models have a rich phenomenology, and can serve as effective descriptions of more complicated theories \cite{Baek:2014jga, Craig:2014lda, Bishara:2015cha, Fedderke:2015txa, arXiv:1506.04149, arXiv:1506.08805, Chen:2015dea, DiFranzo:2015nli, Aravind:2015xst, Ko:2016xwd, Cuoco:2016jqt, Dupuis:2016fda, Das:2016fwl, DiBari:2016guw, arXiv:1611.09675, arXiv:1612.01973, arXiv:1701.08134, arXiv:1704.05359, arXiv:1704.01904, Kolb:2017jvz, Baum:2017enm, Beniwal:2015sdl, Bhattacharya:2016ysw}. They can produce distinct signals at present and future colliders, DM direct detection experiments or in cosmic ray experiments.  In the recent literature, experimental limits on Higgs portal models were considered from Large Hadron Collider (LHC), Circular Electron Positron Collider and Linear Collider searches, LUX and PandaX, supernovae, charged cosmic and gamma rays, Big Bang Nucleosynthesis, and cosmology \cite{Balazs:2014jla, arXiv:1501.05479, Balazs:2015boa, arXiv:1505.04192, arXiv:1506.06556, arXiv:1507.00886, arXiv:1507.01793, arXiv:1507.06158, arXiv:1509.04282, arXiv:1510.06165, arXiv:1512.04119, arXiv:1601.06232, arXiv:1604.04552, arXiv:1604.04589, Assamagan:2016azc, arXiv:1609.03551, arXiv:1609.09079, Cuoco:2017rxb, arXiv:1704.00730, Cai:2017wdu, arXiv:1705.02149, Tu:2017dhl, Fradette:2017sdd, Hoferichter:2017olk, Ellis:2017ple, Dutta:2017sod}. The lack of such signals to date places stringent constraints on Higgs portal models.

The first global study of the scalar Higgs portal DM model was performed in Ref.~\cite{Cheung:2012xb}.  The most recent global fits \cite{SSDM,SSDM2} included relic density constraints from \emph{Planck}, leading direct detection constraints from LUX, XENON1T, PandaX and SuperCDMS, upper limits on the gamma-ray flux from DM annihilation in dwarf spheroidal galaxies with the \textit{Fermi}-LAT, limits on solar DM annihilation from IceCube, and constraints on decays of SM-like Higgs bosons to scalar singlet particles. The most recent \cite{SSDM2} also considered the $\mathbb{Z}_3$ symmetric version of the model, and the impact of requiring vacuum stability and perturbativity up to high energy scales.

In this paper, we perform the first global fits of the effective vector, Majorana fermion and Dirac fermion Higgs portal DM models using the \gambit package \cite{gambit}. By employing the latest data from the DM abundance, indirect and direct DM search limits, and the invisible Higgs width, we systematically explore the model parameter space and present both frequentist and Bayesian results. In our fits, we include the most important SM, nuclear physics, and DM halo model nuisance parameters. For the fermion DM models, we present a Bayesian model comparison between the CP-conserving and CP-violating versions of the theory. We also carry out a model comparison between scalar, vector and fermion DM models.

In Sec.~\ref{sec:models}, we introduce the effective vector and fermion Higgs portal DM models. We describe the constraints that we use in our global fits in Sec.~\ref{sec:constraints}, and the details of our parameter scans in Sec.~\ref{sec:scan-details}. We present likelihood and Bayesian model comparison results respectively in Secs.~\ref{sec:results} and \ref{sec:evidence}, and conclude in Sec.~\ref{sec:conclusions}. Appendix~\ref{sec:DDCalc_feat} documents new features included in the latest version of \ddcalc. Appendix~\ref{sec:cross_sections} contains all the relevant expressions for the DM annihilation rate into SM particles. All \gambit input files, samples and best-fit points for this study are publicly available online via \textsf{Zenodo} \cite{the_gambit_collaboration_2018_1400654}.

%%%%%%%%%%%%%%%%%%%%%%%%%%%%%%%%%%%%%%%%%%%%%%%%%%%%%%%%%%%%%%%%%%%%%%%%%%%%%%%%%%%%%%%%%%%%%%%%%%%%%%%%%%%%%%%%%%%%%%%%
\section{Models}\label{sec:models}
We separately consider vector ($V_\mu$), Majorana fermion ($\chi$) and Dirac fermion ($\psi$) DM particles that are singlets under the SM gauge group. By imposing an unbroken global $\mathbb{Z}_2$ symmetry, under which all SM fields transform trivially but $(V_\mu, \chi, \psi) \rightarrow -(V_\mu, \chi, \psi)$, we ensure that our DM candidates are absolutely stable.

Before electroweak symmetry breaking (EWSB), the Lagrangians for the three different scenarios are \cite{Beniwal:2015sdl}
\begin{align}
  \lagr_{V} &= \lagrsm -\frac{1}{4} W_{\mu\nu} W^{\mu\nu} + \frac{1}{2} \mu_V^2 V_\mu V^\mu - \frac{1}{4!} \lambda_{V} (V_\mu V^\mu)^2 \nonumber \\
  &\hspace{4mm} + \frac{1}{2} \lambda_{hV} V_\mu V^\mu H^\dagger H, \label{eq:Lag_V} \\
  \lagr_{\chi} &= \lagrsm + \frac{1}{2} \ovr{\chi} (i\slashed{\partial} - \mu_\chi) \chi \nonumber \\
  &\hspace{4mm} - \frac{1}{2}\frac{\lambda_{h\chi}}{\Lambda_\chi} \Big(\cos\theta \, \ovr{\chi}\chi + \sin\theta \, \ovr{\chi}i\gamma_5 \chi \Big) H^\dagger H, \label{eq:Lag_chi}\\
  \lagr_{\psi} &= \lagrsm + \ovr{\psi} (i \slashed{\partial} - \mu_\psi) \psi \nonumber \\
  &\hspace{4mm} - \frac{\lambda_{h\psi}}{\Lambda_\psi} \Big(\cos\theta \, \ovr{\psi}\psi + \sin\theta \, \ovr{\psi}i\gamma_5 \psi \Big) H^\dagger H, \label{eq:Lag_psi}
\end{align}
where $\lagrsm$ is the SM Lagrangian, $W_{\mu\nu} \equiv \partial_\mu V_\nu - \partial_\nu V_\mu$ is the vector field strength tensor, $\lambda_{hV}$ is the dimensionless vector Higgs portal coupling, $\lambda_{h\chi,h\psi}/\Lambda_{\chi,\psi}$ are the dimensionful fermionic Higgs portal couplings, and $H$ is the SM Higgs doublet. The fermionic Lagrangians include both CP-odd and CP-even Higgs-portal operators, with $\theta$ controlling their relative size. The choice $\cos\theta = 1$ corresponds to a pure scalar, CP-conserving interaction between the fermionic DM and the SM Higgs field, whereas $\cos\theta = 0$ corresponds to a pure pseudoscalar, maximally CP-violating interaction.  We discuss a possible ultraviolet (UV) completion of such a model in Sec.~\ref{subsec:unitarityEFT} (see also Refs.~\cite{Kim:2008pp,LopezHonorez:2012kv}).

Although all operators in the vector DM model have mass dimension four, the model itself is fundamentally non-renormalisable, as we do not impose a gauge symmetry to forbid e.g.~the mass term for the vector field. Processes with large energies compared to the vector DM mass will violate perturbative unitarity: for large momentum, longitudinal modes of the vector propagator become constant and cross-sections become divergent. In this study we remain agnostic as to the origin of the vector mass term and the quartic vector self-interaction, however we do consider perturbative unitarity in Sec.~\ref{subsec:unitarityEFT}.

After EWSB, the Higgs field acquires a non-zero vacuum expectation value (VEV). In the unitary gauge, we can write
\begin{equation}
  H = \frac{1}{\sqrt{2}}
  \begin{pmatrix}
    0 \\
    v_0 + h
  \end{pmatrix} \, ,
\end{equation}
where $h$ is the physical SM Higgs field and $\vo = (\sqrt{2} G_F)^{-1/2} = 246.22$\,GeV is the Higgs VEV. Thus, the $H^\dagger H$ terms in Eqs.~(\ref{eq:Lag_V}--\ref{eq:Lag_psi}) generate mass and interaction terms for the DM fields. The tree-level physical mass of the vector DM is
\begin{equation}\label{eqn:V-mass}
  m_V^2 = \mu_V^2 + \frac{1}{2}\lambda_{hV} \vo^2 \, .
\end{equation}
For the fermion DM models, the pseudoscalar term (proportional to $\sin\theta$) generates a non-mass-type term that is purely quadratic in the DM fields (e.g., $\ovr{\psi} \gamma_5 \psi$). Therefore after EWSB, to eliminate this term, we perform a chiral rotation of the fermion DM fields through
\begin{equation}
  \chi \rightarrow e^{i\gamma_5 \alpha/2} \chi, \quad \psi \rightarrow e^{i\gamma_5 \alpha/2} \psi \, ,
\end{equation}
where $\alpha$ is a real, space-time independent parameter.\footnote{Note that for the Majorana case, the 4-component spinor can be written in terms of one two-component Weyl spinor. This transformation simply corresponds to a phase transformation of this two-component spinor.} Using the details outlined in the appendix of Ref.~\cite{Beniwal:2015sdl}, we arrive at the following post-EWSB fermion DM Lagrangians
\begin{align}
  \lagr_\chi &= \lagrsm + \frac{1}{2} \ovr{\chi} (i\slashed{\partial} - m_\chi) \chi \nonumber \\
  &\hspace{4mm} - \frac{1}{2}\frac{\lambda_{h\chi}}{\Lambda_\chi} \Big[\cos\xi \, \ovr{\chi} \chi + \sin\xi \, \ovr{\chi} i\gamma_5 \chi \Big] \left(\vo h + \frac{1}{2} h^2 \right) \, , \label{eq:Lag_chi_pEWSB} \\
  \lagr_\psi &= \mathcal{L}_{\textrm{SM}} + \ovr{\psi} (i\slashed{\partial} - m_\psi) \psi \nonumber \\
  &\hspace{4mm} -\frac{\lambda_{h\psi}}{\Lambda_\psi} \Big[\cos\xi \, \ovr{\psi} \psi + \sin\xi \, \ovr{\psi} i\gamma_5 \psi \Big] \left(\vo h + \frac{1}{2} h^2 \right) \, ,
  \label{eq:Lag_psi_pEWSB}
\end{align}
where $\xi \equiv \theta + \alpha$,
\begin{equation}
\cos\xi = \frac{\mu_{\chi,\psi}}{m_{\chi,\psi}} \left(\cos\theta + \frac{1}{2}\frac{\lambda_{h\chi,{h\psi}}}{\Lambda_{\chi,\psi}} \frac{\vo^2}{\mu_{\chi,\psi}} \right),
\end{equation}
and
\begin{align}
m_{\chi,\psi} &= \left[ \left(\mu_{\chi,\psi} + \frac{1}{2}\frac{\lambda_{h\chi,h\psi}}{\Lambda_{\chi,\psi}} \vo^2 \cos\theta \right)^2 \right. \nonumber\\
&\left.\hspace{10.5mm} + \left(\frac{1}{2}\frac{\lambda_{h\chi,h\psi}}{\Lambda_{\chi,\psi}}\vo^2 \sin\theta \right)^2 \right]^{1/2} \, .
\end{align}
In particular, we note that a theory that is CP-conserving before EWSB ($\cos \theta = 1$) is still CP-conserving after EWSB ($\cos \xi = 1$). Because the simplest UV completion leads to $\cos \theta = 1$, this means the particular choice of $\cos \xi = 1$ is also natural from the UV perspective.\footnote{This is not the case for the maximally CP-violating choice $(\cos \theta = 0)$ as EWSB induces a scalar interaction term with $\cos \xi \propto v_0^2$ \cite{Fedderke:2014wda}.} In light of this, we compare the viability of a CP-conserving scenario to the most general case with arbitrary $\xi$ in Sec.~\ref{sec:evidence}.

%%%%%%%%%%%%%%%%%%%%%%%%%%%%%%%%%%%%%%%%%%%%%%%%%%%%%%%%%%%%%%%%%%%%%%%%%%%%%%%%%%%%%%%%%%%%%%%%%%%%%%%%%%%%%%%%%%%%%%%%
\section{Constraints}\label{sec:constraints}

The free parameters of the Lagrangians are subject to various observational and theoretical constraints. For the case of vector DM, the relevant parameters after EWSB are the vector DM mass $m_V$ and the dimensionless coupling $\lambda_{hV}$.\footnote{The quartic self-coupling $\lambda_V$ does not play any role in the DM phenomenology that we consider, and can be ignored. However, it is vital if constraints from electroweak vacuum stability and model perturbativity are imposed \cite{Kahlhoefer:2015jma}. For a global fit including vacuum stability of scalar DM, see e.g., Ref.~\cite{SSDM2}.} The post-EWSB fermion Lagrangians contain three free parameters: the fermion DM mass $m_{\chi,\psi}$, the dimensionful coupling $\lambda_{h\chi,h\psi}/\Lambda_{\chi,\psi}$ between DM and the Higgs, and the scalar-pseudoscalar mixing parameter $\xi$.

In Table~\ref{tab:likelihoods}, we summarise the various likelihoods used to constrain the model parameters in our global fit. In the following subsections, we will discuss both the physics as well as the implementation of each of these constraints.

\begin{table}[t]
  \centering
  \begin{tabular}{ccc}
    \hline
    Likelihoods & \gambit modules/backends & Ref. \\ \hline
    Relic density (\emph{Planck}) & \darkbit & \cite{Ade:2015xua} \\[0.5mm]
    Higgs invisible width & \decaybit & \cite{SDPBit} \\[0.5mm]
    \emph{Fermi}-LAT dSphs & \gamlike \textsf{1.0.0} & \cite{Ackermann:2015zua} \\[0.5mm]
    LUX 2016 (Run II) & \ddcalc \textsf{2.0.0} & \cite{Akerib:2016vxi} \\[0.5mm]
    PandaX 2016 & \ddcalc \textsf{2.0.0} & \cite{Tan:2016zwf} \\[0.5mm]
    PandaX 2017 & \ddcalc \textsf{2.0.0} & \cite{Cui:2017nnn} \\[0.5mm]
    XENON1T 2018 & \ddcalc \textsf{2.0.0} & \cite{Aprile:2018dbl} \\[0.5mm]
    CDMSlite & \ddcalc \textsf{2.0.0} & \cite{Agnese:2015nto} \\[0.5mm]
    CRESST-II & \ddcalc \textsf{2.0.0} & \cite{Angloher:2015ewa} \\[0.5mm]
    PICO-60 2017 & \ddcalc \textsf{2.0.0} & \cite{Amole:2017dex} \\[0.5mm]
    DarkSide-50 2018 & \ddcalc \textsf{2.0.0} & \cite{Agnes:2018fwg} \\[0.5mm]
    IceCube 79-string & \nulike \textsf{1.0.6} & \cite{Aartsen:2012kia} \\
    \hline
  \end{tabular}
  \caption{Likelihoods and corresponding \gambit modules/backends employed in our global fit.}
\label{tab:likelihoods}
\end{table}

\subsection{Thermal relic density}\label{sec:relicdensity}
The time evolution of the DM number density $n_X$ is governed by the Boltzmann equation \cite{Gondolo:1990dk}
\begin{align} \label{eq:Boltzmann}
 \frac{dn_X}{dt} + 3Hn_X = -\langle\sigma v_\textrm{rel}\rangle \left(n_X^2-n_{X,\textrm{eq}}^2\right) \, ,
\end{align}
where $n_{X,\textrm{eq}}$ is the number density at equilibrium, $H$ is the Hubble rate and $\langle\sigma v_{\textrm{{rel}}}\rangle$ is the thermally averaged cross-section times relative (Møller) velocity, given by
\begin{align} \label{eq:sigmavthermal_def}
 \langle\sigma v_\textrm{rel}\rangle = \int^{\infty}_{4m_X^2} ds \, \frac{s\sqrt{s-4m_X^2}K_1\left(\sqrt{s}/T\right)}{16Tm_X^4K_2^2\left(m_X/T\right)} \sigma v_\textrm{rel}^\textrm{cms} \, ,
\end{align}
where $v_\textrm{rel}^\textrm{cms}$ is the relative velocity of the DM particles in the centre-of-mass frame, and $K_{1,2}$ are modified Bessel functions. In the case of non-self-conjugate DM, the right hand side of Eq.~\eqref{eq:Boltzmann} is divided by two.

In the scenarios discussed above, the annihilation process of DM receives contributions from all kinematically accessible final states involving massive SM fields, including neutrinos. Annihilations into SM gauge bosons and fermions are mediated by a Higgs boson in the $s$-channel; consequently, near the resonance region, where $m_X \simeq m_h/2$, it is crucial to perform the actual thermal average as defined in Eq.~(\ref{eq:sigmavthermal_def}) instead of expanding $\sigma v_\textrm{rel}^\textrm{cms}$ into partial waves.\footnote{We assume DM to be in a local thermal equilibrium (LTE) during freeze-out. As pointed out in Ref.~\cite{Binder:2017rgn}, this assumption can break down very close to the resonance, thereby requiring a full numerical solution of the Boltzmann equation in phase space. As this part of the parameter space is in any case very difficult to test experimentally (see Sec.~\ref{sec:results}), we stick to the standard approximation of LTE.} Moreover, we take into account the important contributions arising from the production of off-shell pairs of gauge bosons $W W^*$ and $Z Z^*$~\cite{Cline:2012hg}. To this end, for $45 \;\mathrm{GeV} \leq \sqrt{s} \leq 300 \; \mathrm{GeV}$, we compute the annihilation cross-section into SM gauge bosons and fermions in the narrow-width approximation via
\begin{align} \label{eq:ann_xsecs}
  \sigma v_\textrm{rel}^\mathrm{cms} = P(X)\frac{2\lambda_{hX}^2v_0^2}{\sqrt{s}}\frac{\Gamma_h\left(m_h^*=\sqrt{s}\right)}{\left(s-m_h^2\right)^2+m_h^2\Gamma_h^2\left(m_h\right)} \, ,
\end{align}
where we employ the tabulated Higgs branching ratios $\Gamma(m_h^*)$ as implemented in \decaybit \cite{SDPBit}. For fermionic DM, the dimensionful coupling is implied, $\lambda_{hX} \in \{\lambda_{hV}, \lambda_{h\psi}/\Lambda_\psi, \lambda_{h\chi}/\Lambda_\chi\}$. The pre-factor $P(X)$ is given by
\begin{align} \label{eq:prefactor}
  P(X) =
  \begin{cases}
    \dfrac{1}{9}\left(3-\dfrac{s}{m_V^2} + \dfrac{s^2}{4m_V^4}\right), & X = V_\mu, \\[4mm]
%    \dfrac{s}{2}\left[\left(1-\dfrac{4m_X^2}{s}\right) \cos^2\xi  + \sin^2\xi\right], & X = \psi, \chi.
    \dfrac{s}{2}\left(1-\dfrac{4m_X^2\cos^2\xi}{s}\right), & X = \psi, \chi \, .
  \end{cases}
\end{align}
In particular, we notice that for CP-conserving interactions of a fermionic DM particle, the annihilation cross-section is $p$-wave suppressed.

As shown in Ref.~\cite{Cline:2012hg}, for $\sqrt{s} \gtrsim 300$\,GeV the Higgs 1-loop self-interaction begins to overestimate the tabulated Higgs boson width in Ref.~\cite{Dittmaier:2011ti}. Thus, for $\sqrt{s} > 300$\,GeV (where the off-shell production of gauge boson pairs is irrelevant anyway), we revert to the tree-level expressions for the annihilation processes given in Appendix~\ref{sec:cross_sections}.
Moreover, for $m_X \geq m_h$, DM can annihilate into a pair of Higgs bosons, a process which is not included in Eq.~\eqref{eq:ann_xsecs}. We supplement the cross-sections computed from the tabulated \decaybit values with this process for $m_X \geq m_h$. The corresponding analytical expression for the annihilation cross-sections are given in Appendix~\ref{sec:cross_sections}.

Finally, we obtain the relic density of $X$ by numerically solving Eq.~\eqref{eq:Boltzmann} at each parameter point, using the routines implemented in \darksusy \cite{darksusy,darksusy4} via \darkbit.

In the spirit of the EFT framework employed in this work, we do \emph{not} demand that the particle $X$ constitutes all of the observed DM, i.e., we allow for the possibility  of other DM species to contribute to the observed relic density. Concretely, we implement the relic density constraint using a likelihood that is flat for predicted values below the observed one, and based on a Gaussian likelihood following the \emph{Planck} measured value $\Omega_{\textrm{DM}}h^2 = 0.1188 \pm 0.0010$~\cite{Ade:2015xua} for predictions that exceed the observed central value. We include a $5\%$ theoretical error on the computed values of the relic density, which we combine in quadrature with the observed error on the \emph{Planck} measured value. More details on this prescription can be found in Refs.~\cite{gambit,DarkBit}.

In regions of the model parameter space where the relic abundance of $X$ is less than the observed value, we rescale all predicted direct and indirect detection signals by $f_{\mathrm{rel}} \equiv \Omega_X/\Omega_{\mathrm{DM}}$ and $f_{\mathrm{rel}}^2$, respectively. In doing so, we conservatively assume that the remaining DM population does not contribute to signals in these experiments.

\subsection{Higgs invisible decays}

For $m_X < m_h/2$, the SM Higgs boson can decay into a pair of DM particles, with rates given by~\cite{Beniwal:2015sdl}
\begin{align} \label{eq:invwidth_vector}
  \Gamma_{\mathrm{inv}} (h \rightarrow V V) &= \frac{\lambda_{hV}^2 v_0^2 m_h^3}{128\pi m_V^4} \left(1 - \frac{4 m_V^2}{m_h^2} + \frac{12 m_V^4}{m_h^4} \right) \nonumber \\
  &\hspace{4mm} \times \sqrt{1 - \frac{4 m_V^2}{m_h^2}} \, , \\ \label{eq:invwidth_major}
  \Gamma_{\mathrm{inv}} (h \rightarrow \ovr{\chi} \chi) &= \frac{m_h v_0^2}{16\pi} \left(\frac{\lambda_{h\chi}}{\Lambda_\chi}\right)^2 \left(1 - \frac{4 m_\chi^2 \cos^2 \xi}{m_h^2}\right) \nonumber \\
  &\hspace{4mm} \times \sqrt{1 - \frac{4 m_\chi^2}{m_h^2}} \, , \\ \label{eq:invwidth_dirac}
  \Gamma_{\mathrm{inv}} (h \rightarrow \ovr{\psi} \psi) &= \frac{m_h v_0^2}{8\pi} \left(\frac{\lambda_{h\psi}}{\Lambda_\psi}\right)^2 \left(1 - \frac{4 m_\psi^2 \cos^2 \xi}{m_h^2}\right) \nonumber \\
  &\hspace{4mm} \times \sqrt{1 - \frac{4 m_\psi^2}{m_h^2}},
\end{align}
for the vector, Majorana and Dirac DM scenarios, respectively. These processes contribute to the Higgs invisible width $\Gamma_{\mathrm{inv}}$, which is constrained to be less than 19\% of the total width at $2\sigma$\, C.L.~\cite{Belanger:2013xza}, for SM-like Higgs couplings. We take this constraint into account by using the \decaybit implementation of the Higgs invisible width likelihood, which in turn is based on an interpolation of Fig.~8 in Ref.~\cite{Belanger:2013xza}. Beyond the Higgs invisible width, the LHC provides only a mild constraint on Higgs portal models~\cite{Han:2016gyy}.

\subsection{Indirect detection using gamma rays} \label{sec:gammarays}

Arguably, the most immediate prediction of the thermal freeze-out scenario is that DM particles can annihilate today, most notably in regions of enhanced DM density. In particular, gamma-ray observations of dwarf spheroidal galaxies (dSphs) of the Milky Way are strong and robust probes of any model of thermal DM with unsuppressed annihilation into SM particles.\footnote{We do not include constraints from cosmic-ray antiprotons; although they are potentially competitive with or even stronger than those from gamma-ray observations of dSphs, there is still no consensus on the systematic uncertainty of the upper bound on a DM-induced component in the antiproton spectrum~\cite{Urbano:2014hda, Cuoco:2017iax, Cuoco:2017rxb, Reinert:2017aga}.}

As described in more detail in Ref.~\cite{DarkBit}, the flux of gamma rays in a given energy bin $i$ from a target object labeled by $k$ can be written in the factorised form $\Phi_i \cdot J_k$, where $\Phi_i$ encodes all information about the particle physics properties of the DM annihilation process, while $J_k$ depends on the spatial distribution of DM in the region of interest. For $s$-wave annihilation, one obtains
\begin{align}
 \Phi_i &= \kappa \sum_{j} \frac{(\sigma v)_{0,j}}{8\pi m_X^2}\int_{\Delta E_i} dE \, \frac{dN_{\gamma,j}}{dE} \, , \\
  J_k &= \int_{\Delta\Omega_k} d\Omega \int_{\mathrm{l.o.s.}} ds \, \rho_X^2 \, . \label{eq:def_jk}
\end{align}
Here $\kappa$ is a phase space factor (equal to 1 for self-conjugate DM and 1/2 for non-self-conjugate DM), $(\sigma v)_{0,j}$ is the annihilation cross-section into the final state $j$ in the zero-velocity limit, and $dN_{\gamma,j}/dE$ is the corresponding differential gamma-ray spectrum. The $J$-factor in Eq.~(\ref{eq:def_jk}) is defined via a line of sight (l.o.s.) integral over the square of the DM density $\rho_X$ towards the target object $k$, extended over a solid angle $\Delta\Omega_k$.

In our analysis, we include the \texttt{Pass-8} combined analysis of 15 dwarf galaxies using 6 years of \emph{Fermi}-LAT data~\cite{Ackermann:2015zua}, which currently provides the strongest bounds on the annihilation cross-section of DM into final states containing gamma rays. We use the binned likelihoods implemented in \darkbit~\cite{DarkBit}, which make use of the \gamlike package. Besides the likelihood associated with the gamma-ray observations, given by
\begin{align}
 \ln \lagr_{\rm{exp}} = \sum^{\rm{N_{dSphs}}}_{k=1} \sum^{\rm{N_{eBins}}}_{i=1} \ln \lagr_{ki}\left(\Phi_i \cdot J_k\right) \, ,
\end{align}
we also include a term $\ln \lagr_J$ that parametrises the uncertainties on the $J$-factors \cite{DarkBit,Ackermann:2015zua}. We obtain the overall likelihood by profiling over the $J$-factors of all 15 dwarf galaxies, as
\begin{align}
 \ln \lagr_{\rm{dwarfs}}^{\rm{prof.}} = \underset{J_1\dots J_k}{\textrm{max}}\left(\ln\lagr_{\rm{exp}} + \ln\lagr_J \right) \, .
\end{align}

Let us remark again that for the case of Dirac or Majorana fermion DM with CP-conserving interactions (i.e., $\xi =0$), the annihilation cross-section vanishes in the zero-velocity limit. Scenarios with $\xi \neq 0$ therefore pay the price of an additional penalty from gamma-ray observations, compared to the CP-conserving case.

\subsection{Direct detection} \label{sec:direct_detection}
Direct searches for DM aim to measure the recoil of a nucleus after it has scattered off a DM particle~\cite{Goodman:1984dc}. Following the notation of Ref.~\cite{DarkBit}, we write the predicted number of signal events in a given experiment as
\begin{equation} \label{eq:def_Np}
  N_p = MT_{\mathrm{exp}} \int^{\infty}_0 \phi\left(E\right)\dfrac{dR}{dE} \, dE \, ,
\end{equation}
where $M$ is the detector mass, $T_{\mathrm{exp}}$ is the exposure time and $\phi\left(E\right)$ is the detector efficiency function, i.e., the fraction of recoil events with energy $E$ that are observable after applying all cuts from the corresponding analysis. The differential recoil rate $dR/dE$ for scattering with a target isotope $T$ is given by
\begin{equation} \label{eq:dd_rates}
 \frac{dR}{dE} = \frac{2\rho_0}{m_X} \int v f\left(\bm{v},t\right)\frac{d\sigma}{dq^2}\left(q^2, v\right) \, d^3v \, .
\end{equation}
Here $\rho_0$ is the local DM density, $f(\bm{v},t)$ is the DM velocity distribution in the rest frame of the detector, and $d\sigma/dq^2(q^2, v)$ is the differential scattering cross-section with respect to the momentum transfer $q = \sqrt{2 m_T E}$.

For the vector DM model, the DM-nucleon scattering process is induced by the standard spin-independent (SI) interaction, with a cross-section given by \cite{Beniwal:2015sdl}
\begin{align}
 \sigma_{\mathrm{SI}}^{V} = \frac{\mu_N^2}{\pi}\frac{\lambda_{hV}^2f_N^2 m_N^2}{4m_V^2 m_h^4} \, ,
\end{align}
where $\mu_N = m_V m_N/(m_V + m_N)$ is the DM-nucleon reduced mass and $f_N$ is the effective Higgs-nucleon coupling. The latter is related to the quark content of a proton and neutron, and is subject to (mild) uncertainties. In our analysis we treat the relevant nuclear matrix elements as nuisance parameters; this will be discussed in more detail in Sec.~\ref{subsec:nuisance}.

In the case of fermionic DM $X \in \{\chi,\psi\}$, the pseudoscalar current $\ovr{X} i \gamma_5 X$ induces a non-standard dependence of the differential scattering cross-section on the momentum transfer $q$ (see e.g., Ref.~\cite{Dienes:2013xya}):
\begin{align} \label{eq:fermion_SI_xsec}
 \frac{d\sigma_{\mathrm{SI}}^{X}}{dq^2} &=\frac{1}{v^2} \left(\frac{\lambda_{hX}}{\Lambda_{X}}\right)^2 \frac{A^2 F^2(E) f_N^2m_N^2}{4 \pi m_h^4} \nonumber \\
 &\hspace{4mm} \times \left(\cos^2\xi + \frac{q^2}{4m_X^2}\sin^2\xi\right) \, ,
\end{align}
where $A$ is the mass number of the target isotope of interest, and $F^2(E)$ is the standard form factor for spin-independent scattering~\cite{Lewin:1995rx}. As the typical momentum transfer in a scattering process is $|q| \simeq (1- 100) \; \mathrm{MeV} \ll m_X$, we note that direct detection constraints will be significantly suppressed for scenarios that are dominated by the pseudoscalar interaction, i.e., for $\xi \simeq \pi/2$. For both the vector and fermion models, the spin-dependent (SD) cross-section is absent at leading order. Loop corrections are found not to give a relevant contribution to direct detection in the EFT approach, although they may lead to important effects in specific UV-completions~\cite{Arcadi:2017wqi,Sanderson:2018lmj,Abe:2018emu}.
%\af{In such cases, however, loop corrections in UV complete theories may be important \cite{Sanderson:2018lmj}.} \af{I still think a comment about suppression of SD cross-sections in these models would be useful for the reader.} \Sanjay{Not sure I follow you - there is no SD cross-section in these models at tree-level?}

%\af{Note that the pure pseudoscalar case ($\cos\xi = 0$) is $q^2$ suppressed. The SD DM-nucleon cross-sections, on the other hand, are suppressed in the non-relativistic limit for spin-1/2 and spin-1 DM interacting through a Higgs portal (see e.g., Ref.~\cite{Kumar:2013iva}), and are thus irrelevant.} \Ankit{Andrew, the last sentence above doesn't make sense to me. Do you mean SI DM-nucleon cross is unsuppressed for spin-1 and spin-1/2 $(\cos\xi = 1)$ DM candidates, and thus is relevant?}\af{No. I am commenting on the SD DM-nucleon cross-sections, which are suppressed in these models and thus irrelevant.}

For the evaluation of $N_p$ in Eq.~(\ref{eq:def_Np}), we assume a Maxwell-Boltzmann velocity distribution in the Galactic rest frame, with a peak velocity $v_{\rm{peak}}$ and truncated at the local escape velocity $v_{\rm{esc}}$. We refer to Ref.~\cite{DarkBit} for the conversion to the velocity distribution $f\left(\bm{v},t\right)$ in the detector rest frame. We discuss the likelihoods associated with the uncertainties in the DM velocity distribution in Sec.~\ref{subsec:nuisance}.

We use the \darkbit interface to \ddcalc \textsf{2.0.0}\footnote{\url{http://ddcalc.hepforge.org/}\\ \url{http://github.com/patscott/ddcalc/}} to calculate the number of observed events $o$ in the signal regions for each experiment and to evaluate the standard Poisson likelihood
\begin{align}
 \lagr\left(s|o\right) = \frac{\left(b+s\right)^{o} e^{-(b+s)}}{o!} \, ,
\end{align}
where $s$ and $b$ are the respective numbers of expected signal and background events. We model the detector efficiencies and acceptance rates by
interpolating between the pre-computed tables in \ddcalc.  We include likelihoods from the new XENON1T 2018 analysis \cite{Aprile:2018dbl}, LUX 2016 \cite{Akerib:2016vxi}, PandaX 2016 \cite{Tan:2016zwf} and 2017 \cite{Cui:2017nnn} analyses, CDMSlite \cite{Agnese:2015nto}, CRESST-II \cite{Angloher:2015ewa}, PICO-60 \cite{Amole:2017dex}, and DarkSide-50 \cite{Agnes:2018fwg}. Details of these implementations, as well as an overview of the new features contained in \ddcalc \textsf{2.0.0}, can be found in Appendix~\ref{sec:DDCalc_feat}.

\subsection{Capture and annihilation of DM in the Sun} \label{sec:solarcapture}

Similar to the process underlying direct detection, DM particles from the local halo can also elastically scatter off nuclei in the Sun and become gravitationally bound. The resulting population of DM particles near the core of the Sun can then induce annihilations into high-energy SM particles that subsequently interact with the matter in the solar core.  Of the resulting particles, only neutrinos are able to escape the dense Solar environment. Eventually, these can be detected in neutrino detectors on the Earth~\cite{Press:1985ug,Silk:1985ax,Gould:1987ir}.

The capture rate of DM in the Sun is obtained by integrating the differential scattering cross-section $d\sigma/dq^2$ over the range of recoil energies resulting in a gravitational capture, as well as over the Sun's volume and the DM velocity distribution. To this end, we employ the newly-developed public code \capgen\footnote{\url{https://github.com/aaronvincent/captngen}}, which computes capture rates in the Sun for spin-independent and spin-dependent interactions with general momentum- and velocity-dependence, using the B16 Standard Solar Model \cite{Vinyoles:2016djt} composition and density distribution. We refer to Refs.~\cite{Vincent:2015gqa, Vincent:2016dcp} for details on the capture rate calculation. Notice that similar to direct detection, the capture rate is also subject to uncertainties related to the local density and velocity distribution of DM in the Milky Way. As mentioned earlier, these uncertainties are taken into account by separate nuisance likelihoods to be discussed in Sec.~\ref{subsec:nuisance}.
%
%\begin{equation}
%C(t) = 4\pi \int_0^{R_\odot} dr \, r^2 \int_0^\infty du \, \frac{f_\odot(u)}{u} w \Omega(w),
%\label{caprate}
%\end{equation}
%
%where $u$ is the DM velocity ``at infinity'', while $w = \sqrt{u^2 + v_{\textrm{esc}}(r)^2}$ is the speed at position $r$ including the effect of the Sun's gravitational potential. $f_\odot(u)$ is the DM speed distribution in the frame of the Sun, and $\Omega(w)$ encodes the DM-nucleus elastic scattering cross-section, and is integrated over possible momentum transfers (or equivalently, recoil energies, $E_R$) that lead to scattering below the local escape velocity. In the low-energy limit, if the differential cross-section is velocity-dependent (e.g.\ a p-wave $\implies v^2$-dependent, or d-wave $\implies v^4$-dependent), these powers of $w$ should be included inside the integral over $u$, as detailed in Refs. \cite{Vincent:2015gqa,Vincent:2016dcp}. There is a geometric saturation limit in which all DM that intersects the solar disk is captured, leading to a cutoff on $C(t)$.\footnote{Due to the finite optical depth of the Sun, this is not actually a hard cutoff, see Ref.~\cite{Busoni:2017mhe} for more details.}

Neglecting evaporation (which is well-justified for the DM masses of interest in this study~\cite{Gould:1987ju,Busoni:2013kaa,Busoni:2017mhe}), the total population of DM in the Sun $N_{X}(t)$ follows from
\begin{align} \label{eq:capture_rate}
 \frac{dN_X(t)}{dt} = C(t) - A(t) \, ,
\end{align}
where $C(t)$ is the capture rate of DM in the Sun, and $A(t) \propto \langle \sigma v_\text{rel} \rangle N_X(t)^2$ is the annihilation rate of DM inside the Sun; this is calculated by \darkbit.
We approximate the thermally averaged DM annihilation cross-section, which enters in the expression for the annihilation rate, by evaluating $\sigma v$ at $v=\sqrt{2T_\odot/m_X}$, where $T_\odot = 1.35 \; \mathrm{keV}$ is the core temperature of the Sun.

At sufficiently large $t$, the solution for $N_X(t)$ reaches a steady state and depends only on the capture rate. However, the corresponding time scale $\tau$ for reaching equilibrium depends also on $\sigma v$, and thus changes from point to point in the parameter space. Hence, we use the full solution of Eq.~\eqref{eq:capture_rate} to determine $N_X$ at present times, which in turn determines the normalization of the neutrino flux potentially detectable at Earth. We obtain the flavour and energy distribution of the latter using results from \wimpsim \cite{Blennow:2007tw} included in \darksusy \cite{darksusy,darksusy4}.

Finally, we employ the likelihoods derived from the 79-string IceCube search for high-energy neutrinos from DM annihilation in the Sun \cite{Aartsen:2012kia} using \nulike~\cite{Aartsen:2016exj} via \darkbit; this contains a full unbinned likelihood based on the event-level energy and angular information of the candidate events.

\subsection{Nuisance likelihoods}
\label{subsec:nuisance}

\begin{table}
  \centering
  \begin{tabular}{l@{\,}c@{\,\,}r}
    \hline
    Parameter & & Value ($\pm$Range) \\ \hline
    Local DM density & $\rho_0$ & $0.2$--$0.8$\,GeV\,cm$^{-3}$ \\[1mm]
    Most probable speed & $v_{\rm{peak}}$ & $240\,(24)$\,km s$^{-1}$\\[1mm]
    Galactic escape speed &$v_{\textrm{esc}}$ & $533\,(96)$\,km s$^{-1}$ \\[1mm] \hline
    Nuclear matrix element &$\sigma_s$ & $43\,(24)$\,MeV\\[1mm]
    Nuclear matrix element &$\sigma_l$ & $50\,(45)$\,MeV\\[1mm] \hline
    Higgs pole mass &$\mh$ & $124.1$--$127.3$\,GeV \\[1mm]
    Strong coupling &$\alpha_s^{\MSBar}(m_Z)$ & $0.1181\,(33)$ \\
    \hline
  \end{tabular}
  \caption{Nuisance parameters that are varied simultaneously with the DM model parameters in our scans. All parameters have flat priors. For more details about the nuisance likelihoods, see Sec.~\ref{subsec:nuisance}.}
  \label{tab:nuis_params}
\end{table}

The constraints discussed in the previous sections often depend on \emph{nuisance parameters}, i.e.~parameters not of direct interest but required as input for other calculations. Examples are nuclear matrix elements related to the DM direct detection process, the distribution of DM in the Milky Way, or SM parameters known only to finite accuracy. It is one of the great virtues of a global fit that such uncertainties can be taken into account in a fully consistent way, namely by introducing new free parameters into the fit and constraining them by new likelihood terms that characterise their uncertainty. We list the nuisance parameters included in our analysis in Table~\ref{tab:nuis_params}, and discuss each of them in more detail in the rest of this section.

Following the default treatment in \darkbit, we include a nuisance likelihood for the local DM density $\rho_0$ given by a log-normal distribution with central value $\rho_0 = 0.40$\,GeV\,cm$^{-3}$ and an error $\sigma_{\rho_0}=0.15$\,GeV\,cm$^{-3}$. To reflect the log-normal distribution, we scan over an asymmetric range for $\rho_0$. For more details, see Ref.~\cite{DarkBit}.

For the parameters determining the Maxwell-Boltzmann distribution of the DM velocity in the Milky Way, namely $v_\text{peak}$ and $v_\text{esc}$, we employ simple Gaussian likelihoods. Since $v_{\rm{peak}}$ is equal to the circular rotation speed $v_{\rm{rot}}$ at the position of the Sun for an isothermal DM halo, we use the determination of $v_{\rm{rot}}$ from Ref.~\cite{Reid:2014boa} to obtain $v_{\rm{peak}} = 240\pm8$\,km\,s$^{-1}$.\footnote{Ref.~\cite{Bovy:2012ba} argues that the peculiar velocity of the Sun is somewhat larger than the canonical value ${\bf{v}}_{\odot,\text{pec}} = (11, 12, 7)$\,km\,s$^{-1}$~\cite{Schoenrich:2009bx}, leading to $v_{\rm{rot}} = 218\pm6$\,km\,s$^{-1}$. In the present study we do not consider uncertainties in ${\bf{v}}_{\odot,\text{pec}}$ and therefore adopt the measurement of $v_{\rm{rot}}$ from Ref.~\cite{Reid:2014boa}.} The escape velocity takes a central value of $v_{\rm{esc}} = 533 \pm 31.9$\,km\,s$^{-1}$, where we convert the 90\% C.L.~interval obtained by the RAVE collaboration~\cite{Piffl:2013mla}, assuming that the error is Gaussian.

As noted already in Sec.~\ref{sec:direct_detection}, the scattering cross-section of DM with nuclei (which enters both the direct detection and solar capture calculations) depends on the effective DM-nucleon coupling $f_N$, which is given by~\cite{DarkBit}
\begin{align}
  f_N = \frac{2}{9} + \frac{7}{9}\sum_{q=u,d,s} f_{Tq}^{(N)} \, .
\end{align}
Here $f_{Tq}^{(N)}$ are the nuclear matrix elements associated with the quark $q$ content of a nucleon $N$. As described in more detail in Ref.~\cite{Cline:2013gha}, these are obtained from the following observable combinations
\begin{equation}
  \sigma_l \equiv m_l \langle N | \ovr{u} u + \ovr{d} d | N \rangle, \quad \sigma_s \equiv m_s \langle N | \ovr{s} s | N \rangle \, ,
\end{equation}
where $m_l \equiv (m_u + m_d)/2$. We take into account the uncertainty on these matrix elements via Gaussian likelihoods given by $\sigma_s = 43 \pm 8$\,MeV~\cite{Lin:2011ab} and $\sigma_l = 50 \pm 15$\,MeV~\cite{Bishara:2017pfq}. The latter deviates from the default choice implemented in \darkbit as it reflects recent lattice results, which point towards smaller values of $\sigma_l$ (see Ref.~\cite{Bishara:2017pfq} for more details). Furthermore, we have confirmed that the uncertainties on the light quark masses have a negligible impact on $f_N$. Thus, for simplicity, we do not include them in our fit.

We also use a Gaussian likelihood for the Higgs mass, based on the PDG value of $m_h = 125.09\pm0.24$\,GeV \cite{Patrignani:2016xqp}. In line with our previous study of scalar singlet DM \cite{SSDM}, we allow the Higgs mass to vary by more than $4\sigma$  as the phenomenology of our models depends strongly on $m_h$, most notably near the Higgs resonance region. Finally, we take into account the uncertainty on the strong coupling constant $\alpha_s$, which enters the expression for the DM annihilation cross-section into SM quarks (see Appendix~\ref{sec:cross_sections}), taking a central value $\alpha_s^{\MSBar}(m_Z) = 0.1181 \pm 0.0011$ \cite{Patrignani:2016xqp}.

\subsection{Perturbative unitarity and EFT validity}
\label{subsec:unitarityEFT}

The parameter spaces in which we are interested are limited by the requirement of perturbative unitarity. First of all, this requirement imposes a bound on any dimensionless coupling in the theory. Furthermore, as neither the vector or fermion Higgs portal models are renormalisable, we must ensure that the effective description is valid for the parameter regions to be studied.

The dimensionless coupling $\lambda_{hV}$ in the vector DM model is constrained by the requirement that annihilation processes such as $VV\to hh$ do not violate perturbative unitarity. Determining the precise bound to be imposed on $\lambda_{hV}$ is somewhat involved, so we adopt the rather generous requirement $\lambda_{hV} < 10$ with the implicit understanding that perturbativity may become an issue already for somewhat smaller couplings.

For small DM masses, an additional complication arises from the fact that theories with massive vector bosons are not generally renormalisable. In that case cross-sections do not generally remain finite in the $m_V \to 0$ limit and a significant portion of parameter space violates perturbative unitarity ~\cite{Lebedev:2011iq}. However, by restricting ourselves to the case of $\mu^2_V, \lambda_{hV} \geq 0$ we can safely tackle both issues due to the fact that $m_V \to 0$ implies $\lambda_{hV} \to 0$. Using Eq.~\eqref{eqn:V-mass}, this condition translates to
\begin{equation} \label{eq:pert_unitarity}
 0 \leq \lambda_{hV} \leq \frac{2 m_V^2}{v_0^2} \, .
\end{equation}
A more careful analysis might lead to a slightly larger valid parameter space, but as we will see in Sec.~\ref{sec:vector_model_results}, those regions would be excluded by the Higgs invisible width anyway.

The EFT validity of the fermion DM models depends on the specific UV completion. To estimate the range of validity, we consider a UV completion in which a heavy scalar mediator field $\Phi$
couples to the fermion DM $X$ and the Higgs doublet as \cite{Kim:2008pp}
\begin{align}
 \lagr \supset - \mu g_H \Phi H^\dagger H - g_X \Phi \overline{X}\left(\cos\theta + i\sin\theta\gamma_5\right)X \, ,
\end{align}
where $X \in \{\chi,\psi\}$ and $\mu$ has mass dimension 1.\footnote{Note that the $\gamma_5$ term can be generated by a complex mass term $\widetilde{m}_X$ in the original fermion Lagrangian and performing a chiral rotation. Thus, full CP conservation $(\cos\theta = 1)$ is equivalent to having a real mass term.} For this specific UV completion, we assume that the mixing between $\Phi$ and the Higgs field is negligible and can be ignored. The heavy scalar field can be integrated out to give a dimensionful coupling in the EFT approximation as
\begin{align} \label{eq:uv_fermion}
 \lagr \supset - \frac{\mu g_X g_H}{m_\Phi^2} H^\dagger H \overline{X}\left(\cos\theta + i\sin\theta\gamma_5\right)X \, .
\end{align}
Thus, by comparing Eq.~\eqref{eq:uv_fermion} with the fermion DM Lagrangians in Eqs.~\eqref{eq:Lag_chi} and \eqref{eq:Lag_psi}, we can identify $\mu g_X g_H/m_\Phi^2$ with $\lambda_{hX}/\Lambda_X$. As $\mu$ should be set by the new physics scale, we take it to be roughly $m_\Phi$, implying $g_X g_H /m_\Phi \sim \lambda_{hX}/\Lambda_X$. In addition, we require the couplings to be perturbative, i.e., $g_X g_H \leq 4\pi$.

We need to consider the viable scales for which this approximation is valid. We require that the mediator mass $m_\Phi$ is far greater than the momentum exchange $q$ of the interaction, i.e., $m_\Phi \gg q$ such that $\Phi$ can be integrated out. For DM annihilations, the momentum exchange is $q \approx 2m_X$. Thus, the EFT approximation breaks down when $m_\Phi < 2m_X$ and our EFT assumption is violated when
\begin{equation}
  \frac{\lambda_{hX}}{\Lambda_{X}} \geq \frac{4\pi}{2 m_X} \, .
\end{equation}
As the typical momentum transfer in a direct detection experiment is roughly on the order of a few MeVs, the EFT validity limit requires $m_\Phi \gg \mathcal{O}\left(\mathrm{MeV}\right)$, which is always satisfied by the previous demand $m_\Phi > 2m_X$ for the mass ranges of interest. In this case, we assume that the couplings saturate the bound from perturbativity, i.e., $g_X g_H = 4\pi$; the constraint would be stronger if the couplings were weaker.

For parameter points close to the EFT validity bound, the scale of new physics is expected to be close to or even below $2m_\chi$. In this case, the annihilation cross-section $\sigma v_\textrm{rel}$, used in predictions of both the relic density and indirect detection signals, may receive substantial corrections from interactions with $\Phi$, which are not captured in the EFT approach. The likelihoods computed for these points should hence be interpreted with care.

Note that this prescription is only the simplest and most conservative approach; additional constraints can be obtained by unitarising the theory (e.g.\ \cite{2016JHEP...08..125B}).

%%%%%%%%%%%%%%%%%%%%%%%%%%%%%%%%%%%%%%%%%%%%%%%%%%%%%%%%%%%%%%%%%%%%%%%%%%%%%%%%%%%%%%%%%%%%%%%%%%%%%%%%%%%%%%%%%%%%%%%%
\section{Scan details}\label{sec:scan-details}

We investigate the Higgs portal models using both Bayesian and frequentist statistics.  The parameter ranges and priors that we employ in our scans of the vector and fermion DM models are summarised in Tables~\ref{tab:vector_params} and \ref{tab:fermi_params}, respectively. Whilst the likelihoods described in the previous sections are a common ingredient in both our frequentist and Bayesian analyses, the priors only directly impact our Bayesian analyses. We discuss our choice of priors in Sec.\ \ref{posteriors}. For a review of our statistical approaches to parameter inference, see e.g., Ref.~\cite{gambit}.

\begin{table}[tbp]
  \centering
  \begin{tabular}{lccc}
    \hline
    Parameter       & Minimum   & Maximum   & Prior type \\ \hline
    $\lambda_{hV}$      & $10^{-4}$   & 10    & log \\[1mm]
    $m_V$ (low mass)        & 45\,GeV   & 70\,GeV   & flat \\[1mm]
    $m_V$ (high mass)       & 45\,GeV   & 10\,TeV   & log  \\ \hline
  \end{tabular}
  \caption{Parameter ranges and priors for the vector DM model.}
  \label{tab:vector_params}
\end{table}
\begin{table}[tbp]
  \centering
  \begin{tabular}{lccc}
    \hline
    Parameter           & Minimum     & Maximum   & Prior type \\ \hline
    $\lambda_{h\chi,h\psi}/\Lambda_{\chi,\psi}$   & $10^{-6}$ GeV$^{-1}$  & 1 GeV$^{-1}$  & log \\[1mm]
    $\xi$             & $0$       & $\pi$   & flat \\[1mm]
    $m_{\chi,\psi}$ (low mass)      & 45\,GeV     & 70\,GeV   & flat \\[1mm]
    $m_{\chi,\psi}$ (high mass)       & 45\,GeV     & 10\,TeV   & log \\ \hline
  \end{tabular}
  \caption{Parameter ranges and priors for the fermion DM models. Our choice for the range of $\xi$ between $0$ and $\pi$ reflects the fact that only odd powers of $\cos\xi$ appear in the observables that we consider, but never odd powers of $\sin\xi$, which cancel exactly due to the complex conjugation. Thus, the underlying physics is symmetric under $\xi \rightarrow -\xi$.}
  \label{tab:fermi_params}
\end{table}

There are two main objectives for the Bayesian scans: firstly, producing marginal posteriors for the parameters of interest, where we integrate over all unplotted parameters, and secondly, computing the marginal likelihood (or Bayesian evidence). We discuss the marginal likelihood in Sec.~\ref{sec:evidence}. We use \twalk, an ensemble Markov Chain Monte Carlo (MCMC) algorithm, for sampling from the posterior, and \textsf{MultiNest}~\cite{Feroz:2007kg,Feroz:2008xx,2013arXiv1306.2144F}, a nested sampling algorithm, for calculating the marginal likelihood. We use \twalk for obtaining the marginal posterior due to the ellipsoidal bias commonly seen in posteriors computed with \multinest \cite{ScannerBit}.

For the frequentist analysis, we are interested in mapping out the highest likelihood regions of our parameter space. For this analysis we largely use \diver, a differential evolution sampler, efficient for finding and exploring the maxima of a multi-dimensional function. Details of \twalk and \diver can be found in Ref.~\cite{ScannerBit}.

Due to the resonant enhancement of the DM annihilation rate by $s$-channel Higgs exchange at $m_X \approx m_h/2$, there is a large range of allowed DM-Higgs couplings that do not overproduce the observed DM abundance. When scanning over the full mass range, it is difficult to sample this resonance region well, especially with a large number of nuisance parameters. For this reason, we perform separate, specific scans in the low-mass region around the resonance, using both \twalk and \diver. When plotting the profile likelihoods, we combine the samples from the low- and high-mass scans.

In addition, as part of the Bayesian analysis, we perform targeted \twalk and \multinest scans of the fermion DM parameter space where the interaction is wholly scalar ($\xi=0$), using the same priors for the fermion DM mass and its dimensionful coupling as in Table~\ref{tab:fermi_params}. This allows us to perform model comparison between the cases where $\xi$ is fixed at zero, and where it is left as a free parameter.

\begin{table}[tbp]
  \centering
  \begin{tabular}{lll}
    \hline
    Scanner   & Parameter     & Value       \\ \hline
    \twalk    & \cpp{chain_number}  & 1370\,(1) \\[1mm]
        & \cpp{sqrtR} $-$ 1 & $<0.01$   \\[1mm]
        & \cpp{timeout_mins} & 1380 \\ \hline
    \multinest  & \cpp{nlive}     & 20,000      \\[1mm]
        & \cpp{tol}   & $10^{-2}$     \\ \hline
    \diver    & \cpp{NP}    & 50,000      \\[1mm]
        & \cpp{convthresh}  & $10^{-5}$     \\ \hline
  \end{tabular}
  \caption{Conversion criteria used for various scanning algorithms in both the full and low mass regimes. The \cpp{chain_number} chosen for \twalk varies from scan to scan; we use the default \twalk behaviour of \cpp{chain_number} = $N_{\text{MPI}}$ + $N_{\text{params}}$ + 1 on 1360 MPI processes. For more details, see Ref.~\cite{ScannerBit}.}
  \label{tab:conv_criteria}
\end{table}

The convergence criteria that we employ for the different samplers are outlined in Table~\ref{tab:conv_criteria}. We carried out all \diver scans on 340 Intel Xeon Phi 7250 (Knights Landing) cores. As in our recent study of scalar singlet DM \cite{SSDM2}, we ran \twalk scans on 1360 cores for 23 hours, providing us with reliable sampling. The \multinest scans are based on runs using 240 Intel Broadwell cores, with a relatively high tolerance value, which is nevertheless sufficient to compute the marginal likelihood to the accuracy required for model comparison. We use the importance sampling log-evidence from \multinest to compute Bayes factors.

\begin{figure*}[tbp]
  \centering
  \includegraphics[height=0.8\columnwidth]{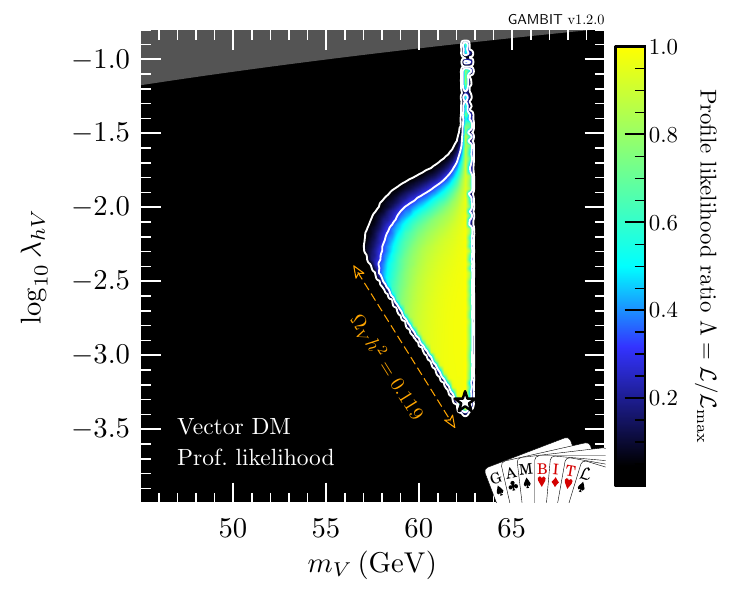}
  \includegraphics[height=0.8\columnwidth]{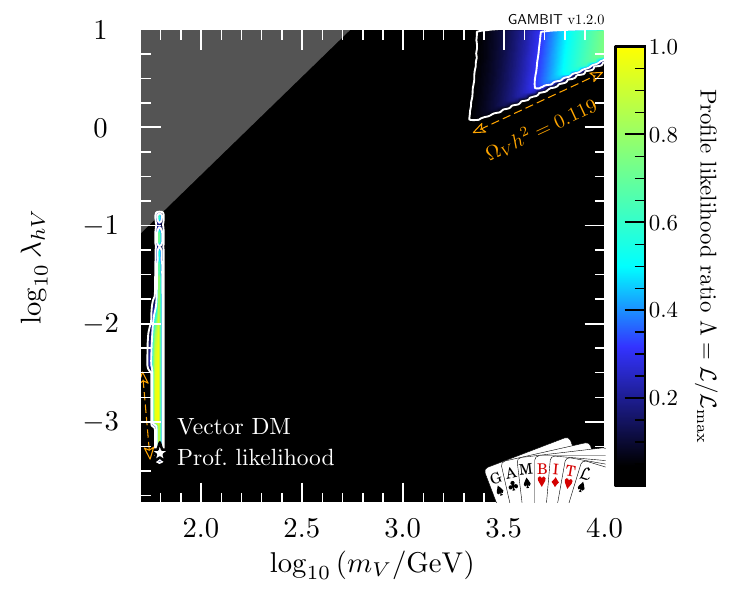}
  \vspace{-2mm}
  \caption{Profile likelihood in the $(m_V,\lambda_{hV})$ plane for vector DM. Contour lines show the $1$ and $2\sigma$ confidence regions. The left panel gives an enhanced view
  of the resonance region around $m_V \sim m_h/2$. The right panel shows the full parameter space explored in our fits. The greyed out region shows points that do not satisfy Eq.~\eqref{eq:pert_unitarity}, the white star shows the best-fit point, and the edges of the preferred parameter space along which the model reproduces the entire observed relic density are indicated with orange annotations.}
  \label{fig::vector_profile}
\end{figure*}

\begin{table*}[tbp]
\center{
  \begin{tabular}{l r c c c c c c}
  & & \multicolumn{6}{c}{ $\Delta\ln\mathcal{L}$}\\
  Log-likelihood contribution                 & Ideal       & $V_\mu$ & $V_\mu$ + RD & $\chi$ & $\chi$ + RD & $\psi$ & $\psi$ + RD\\
  \hline
  Relic density                               & 5.989       & 0.000 & 0.106 & 0.000 & 0.107 & 0.000 & 0.242 \\
  Higgs invisible width                       & 0.000       & 0.000 & 0.000 & 0.000 & 0.001 & 0.000 & 0.000 \\
  $\gamma$ rays (\textit{Fermi}-LAT dwarfs)   & $-$33.244   & 0.105 & 0.105 & 0.102 & 0.120 & 0.129 & 0.134 \\
  LUX 2016 (Run II)                           & $-$1.467    & 0.003 & 0.003 & 0.020 & 0.000 & 0.028 & 0.028 \\
  PandaX 2016                                 & $-$1.886    & 0.002 & 0.002 & 0.013 & 0.000 & 0.018 & 0.017 \\
  PandaX 2017                                 & $-$1.550    & 0.004 & 0.004 & 0.028 & 0.000 & 0.039 & 0.039 \\
  XENON1T 2018                                & $-$3.440    & 0.208 & 0.208 & 0.143 & 0.211 & 0.087 & 0.087 \\
  CDMSlite                                    & $-$16.678   & 0.000 & 0.000 & 0.000 & 0.000 & 0.000 & 0.000 \\
  CRESST-II                                   & $-$27.224   & 0.000 & 0.000 & 0.000 & 0.000 & 0.000 & 0.000 \\
  PICO-60 2017                                & 0.000       & 0.000 & 0.000 & 0.000 & 0.000 & 0.000 & 0.000 \\
  DarkSide-50 2018                            & $-$0.090    & 0.000 & 0.000 & 0.002 & 0.000 & 0.005 & 0.005 \\
  IceCube 79-string                           & 0.000       & 0.000 & 0.000 & 0.000 & 0.000 & 0.001 & 0.001 \\
  Hadronic elements $\sigma_{s}$, $\sigma_l$  & $-$6.625    & 0.000 & 0.000 & 0.000 & 0.000 & 0.000 & 0.000 \\
  Local DM density $\rho_0$                   & 1.142       & 0.000 & 0.000 & 0.000 & 0.000 & 0.000 & 0.000 \\
  Most probable DM speed $v_\mathrm{peak}$    & $-$2.998    & 0.000 & 0.000 & 0.000 & 0.000 & 0.000 & 0.000 \\
  Galactic escape speed $v_\mathrm{esc}$      & $-$4.382    & 0.000 & 0.000 & 0.000 & 0.000 & 0.000 & 0.000 \\
  $\alpha_{\text{s}}$                         & 5.894       & 0.000 & 0.000 & 0.000 & 0.000 & 0.000 & 0.000 \\
  Higgs mass                                  & 0.508       & 0.000 & 0.000 & 0.000 & 0.000 & 0.000 & 0.000 \\ \hline
  Total                                       & 86.051      & 0.322 & 0.428 & 0.308 & 0.439 & 0.307 & 0.553 \\
  \end{tabular}
  \caption{Contributions to the delta log-likelihood $(\Delta \ln \mathcal{L})$ at the best-fit point for the vector, Majorana and Dirac DM, compared to an `ideal' case, both with and without the requirement of saturating the observed relic density (RD). Here `ideal' is defined as the central observed value for detections, and the background-only likelihood for exclusions.  Note that many likelihoods are dimensionful, so their absolute values are less meaningful than any offset with respect to another point (for more details, see Sec.\ 8.3 of Ref.\ \cite{gambit}).}
  \label{tab::bestfit}
}
\end{table*}

For profile likelihood plots, we combine the samples from all \diver and \twalk scans, for each model. The plots are based on $1.46\times10^7$, $1.70\times10^7$ and $1.73\times10^7$ samples for the vector, Majorana and Dirac models, respectively. We do all marginalisation, profiling and plotting with \pippi \cite{Scott:2012qh}.

%%%%%%%%%%%%%%%%%%%%%%%%%%%%%%%%%%%%%%%%%%%%%%%%%%%%%%%%%%%%%%%%%%%%%%%%%%%%%%%%%%%%%%%%%%%%%%%%%%%%%%%%%%%%%%%%%%%%%%%%
\section{Results}\label{sec:results}

\subsection{Profile likelihoods}

In this section, we present profile likelihoods from the combination of all \diver and \twalk scans for the vector, Majorana and Dirac models. Profile likelihoods in the vector model parameters are shown in Fig.~\ref{fig::vector_profile}, with key observables rescaled to the predicted DM relic abundance in Fig.~\ref{fig::vector_rescaled_obs}. Majorana model parameter profile likelihoods are shown in Figs.~\ref{fig::majorana_profile} and ~\ref{fig::majorana_mixing}, with observables in Fig.~\ref{fig::majorana_rescaled_obs}. For the Dirac model, we simply show the mass-coupling plane in Fig.~\ref{fig::dirac_profile}, as the relevant physics and results are virtually identical to the Majorana case.

\subsubsection{Vector model}\label{sec:vector_model_results}

\begin{figure}[tbp]
  \centering
  \includegraphics[height=0.76\columnwidth]{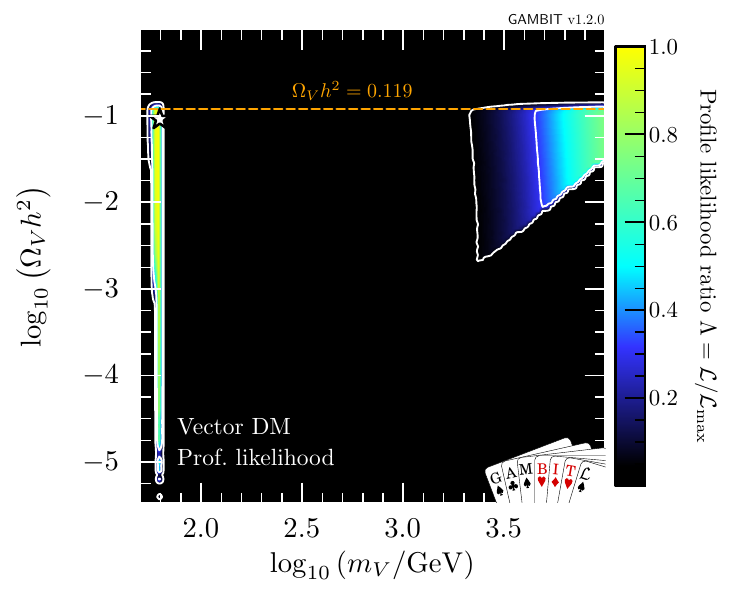}
  \includegraphics[height=0.76\columnwidth]{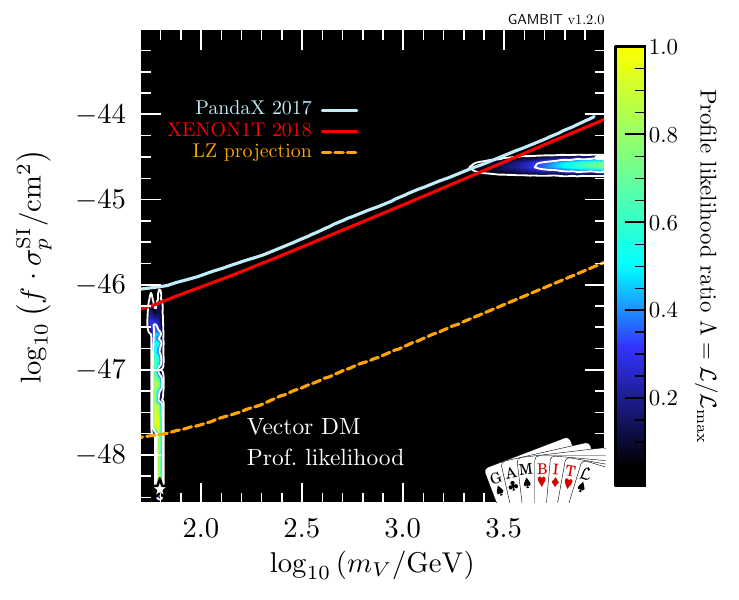}
  \includegraphics[height=0.76\columnwidth]{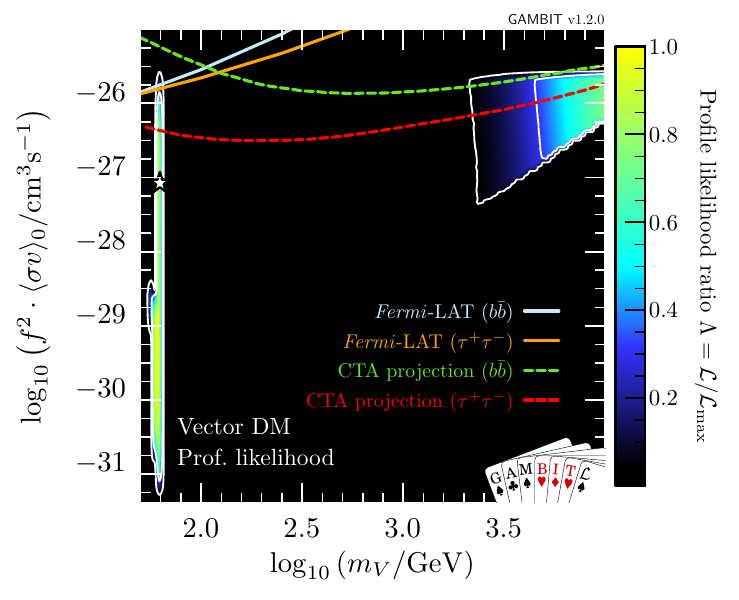}
  \caption{Profile likelihoods for vector DM in planes of observable quantities. \textit{Top:} relic density. \textit{Centre:} spin-independent WIMP-proton cross-section, where solid lines show exclusions from PandaX 2017~\cite{Cui:2017nnn} and XENON1T 2018~\cite{Aprile:2018dbl}, and the dashed line shows the expected sensitivity of LZ \cite{LZ}. \textit{Bottom:} present-day DM annihilation cross-section, where solid lines show published exclusions from \emph{Fermi}-LAT~\cite{Ackermann:2015zua}, and dashed lines show projections from CTA~\cite{Acharya:2017ttl} (see footnote \ref{CTA} for more details). Contour lines in each panel show the $1$ and $2\sigma$ confidence regions, while the white star shows the best-fit point. Cross-sections are rescaled by the fraction of predicted relic abundance $f\equiv\Omega_V/\Omega_{\mathrm{DM}}$.}
  \label{fig::vector_rescaled_obs}
\end{figure}

Fig.~\ref{fig::vector_profile} shows that the resonance region is tightly constrained by the Higgs invisible width from the upper-left when $m_V < m_h/2$, by the relic density constraint from below, and by direct and indirect detection from the right. Nevertheless, the resonant enhancement of the DM annihilation at around $m_h/2$, combined with the fact that we allow for scenarios where $V_\mu$ is only a fraction of the observed DM, permits a wide range of portal couplings. Interestingly, the perturbative unitarity constraint (shown as dark grey) in Eq.~\eqref{eq:pert_unitarity} significantly shortens the degenerate `neck' region that appears exactly at $m_h/2$. Most notably, this is in contrast with the scalar Higgs portal model \cite{SSDM,SSDM2} where no such constraint exists.

The high-mass region contains a set of solutions at $m_V \simeq 10\,\text{TeV}$ and $\lambda_{hV} \gtrsim 1$, which are constrained by the relic density from below and direct detection from the left. This second island is prominent in both our previous studies of the scalar Higgs portal model \cite{SSDM,SSDM2} as well as other studies of the vector Higgs portal \cite{Beniwal:2015sdl}. The precise extent of this region depends on the upper bound imposed on $\lambda_{hV}$ to reflect the breakdown of perturbativity. While the constraint that we apply ensures that perturbative unitarity is not violated~\cite{Lebedev:2011iq}, higher-order corrections may nevertheless become important in this region. The perturbative unitarity constraint from Eq.~\eqref{eq:pert_unitarity} excludes solutions that would otherwise exist in a separate triangular region at $m_\chi \simeq m_h$, $\lambda_{hV}\simeq 1$.

In Table~\ref{tab::bestfit}, we show a breakdown of the contributions to the likelihood at the best-fit point, which lies on the lower end of the resonance region at $\lambda_{hV} = 4.9\times 10^{-4}$ and $m_V = 62.46$\,GeV.  If we demand that vector singlet DM constitutes all of the observed DM, by requiring $\Omega_V h^2$ to be within $1\sigma$ of the observed \emph{Planck} relic abundance, the best-fit point remains almost unchanged, at $\lambda_{hV} = 4.5\times 10^{-4}$ and $m_V = 62.46$\,GeV.  We give details of these best-fit points, along with the equivalent for fermion models, in Table \ref{tab::bestfit_params}.

\begin{figure*}[tbp]
  \centering
  \includegraphics[height=0.8\columnwidth]{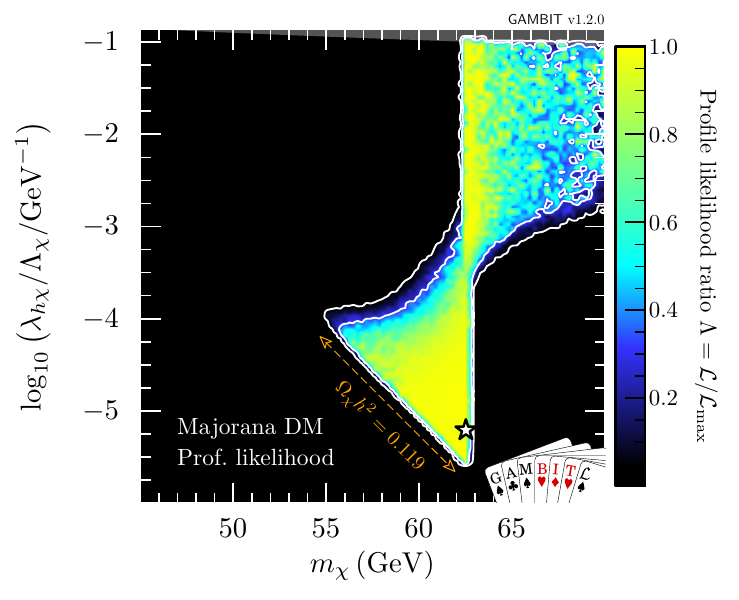}
  \includegraphics[height=0.8\columnwidth]{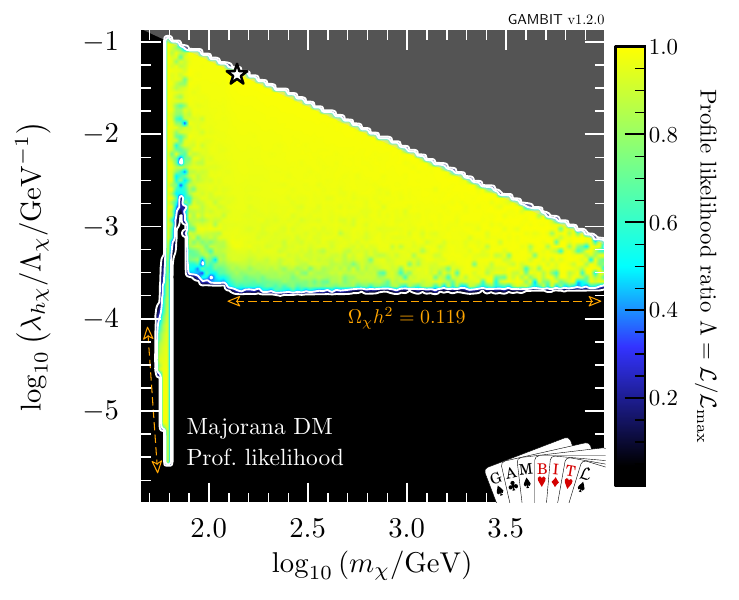}
  \vspace{-2mm}
  \caption{Profile likelihood in the $(m_{\chi}, \lambda_{h\chi}/\Lambda_\chi)$ plane for Majorana fermion DM. Contour lines show the $1$ and $2\sigma$ confidence regions. The left panel gives an enhanced view of the resonance region around $m_{\chi} \sim m_h/2$. The right panel shows the full parameter space explored in our fits. The greyed out region shows where our approximate bound on the validity of the EFT is violated, white stars show the best-fit point for each mass region, and the edges of the preferred parameter space along which the model reproduces the entire observed relic density are indicated with orange annotations.}
  \label{fig::majorana_profile}
\end{figure*}

\begin{figure*}[tbp]
  \centering
  \includegraphics[height=0.8\columnwidth]{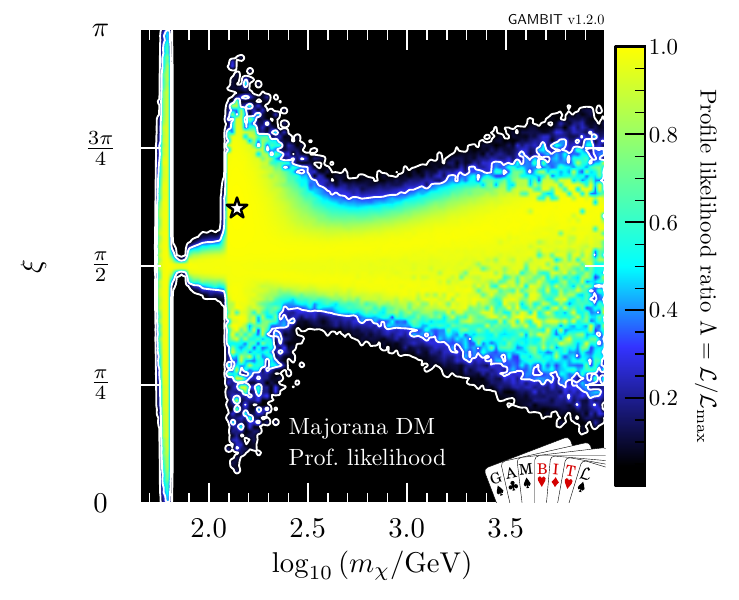}
  \includegraphics[height=0.8\columnwidth]{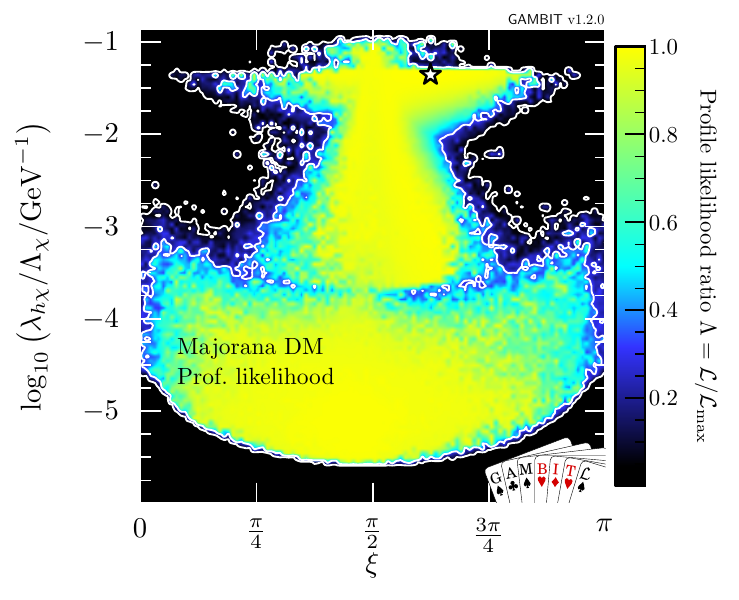}
  \vspace{-2mm}
  \caption{Profile likelihood in the $(m_{\chi},\xi)$ and $(\xi,\lambda_{h\chi}/\Lambda_\chi)$ planes of the Majorana fermion model. Contour lines show the $1$ and $2\sigma$ confidence regions. The white star shows the best-fit point.}
  \label{fig::majorana_mixing}
\end{figure*}

In Fig.~\ref{fig::vector_rescaled_obs}, we show the relic density of the vector model (top), as well as the cross-sections relevant for direct (centre) and indirect detection (bottom), all plotted as a function of mass. Only models along the lower-$\lambda_{hV}$ edge of the two likelihood modes have relic densities equal to the observed value. Larger values of $\lambda_{hV}$ result in progressively larger annihilation cross-sections and therefore more suppression of the relic density, cancelling the corresponding increase in $\sigma^{\textrm{SI}}_p$ and resulting in an essentially constant rescaled cross-section $f \cdot \sigma^{\textrm{SI}}_p \sim 10^{-45}\,\text{cm}^{-2}$ in the remaining allowed high-mass region. Future direct detection experiments such as LZ \cite{LZ} will be able to probe the high-mass region in its entirety. However, the best-fit point -- near the bottom of the resonance region -- will remain out of reach. Future indirect searches, such as the Cherenkov Telescope Array (CTA)\footnote{\label{CTA}The CTA projections plotted in Fig.~\ref{fig::vector_rescaled_obs} assume an Einasto density profile, and are based on 500 hours of observations of the Galactic centre~\cite{Acharya:2017ttl}, with no systematic uncertainties. They should therefore be considered optimistic \cite{Silverwood:2014yza,Pierre14}. }~\cite{Acharya:2017ttl} will also be able to probe large amounts of the high-mass region; however it does not have the exclusion power that direct detection does for Higgs portal models. Again, the best-fit point remains out of reach.

\subsubsection{Majorana fermion model}

We show profile likelihoods in the $(m_{\chi},\lambda_{h\chi}/\Lambda_\chi)$ plane in Fig.~\ref{fig::majorana_profile}, with the low-mass region in the left panel and the full mass region in the right panel. Here, there are no longer two distinct solutions: the resonance and high mass regions are connected. From the left panel in Fig.~\ref{fig::majorana_mixing}, where we plot the profile likelihood in the $(m_{\chi},\xi)$ plane, we can see that these regions are connected by the case where the portal interaction is purely pseudoscalar, $\xi = \pi/2$, leading to an almost complete suppression of constraints from the direct detection experiments, as given in Eq.\ \eqref{eq:fermion_SI_xsec}.

The high mass region prefers $\xi \sim \pi / 2$, with a wider deviation from $\pi/2$ permitted as $m_\chi$ is increased, due to direct detection constraints, which become less constraining at higher WIMP masses. There is an enhancement in the permitted range of mixing angles at $m_\chi \gtrsim m_h$, due to the contact term $(\propto \ovr{\chi}\chi hh)$, where DM annihilation to on-shell Higgses reduces the relic density, providing another mechanism for suppressing direct detection signals, thus lifting the need to tune $\xi$.

The results are roughly symmetric about $\xi=\pi/2$, however due to odd powers of $\cos\xi$ in the annihilation cross-section (see Appendix ~\ref{sec:cross_sections}), there is a slight asymmetry for masses above $m_h$. This is most clearly seen in the triangular `wings' at $m_\chi\gtrsim m_h$ in Fig.~\ref{fig::majorana_mixing} where there are more solutions for $\xi > \pi/2$ than for $\xi < \pi/2$.

\begin{figure}[tbp]
  \centering
  \includegraphics[height=0.8\columnwidth]{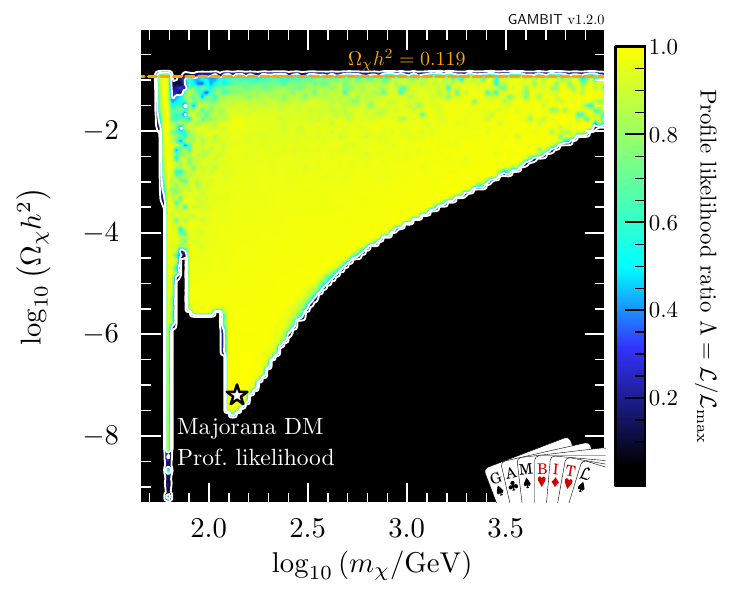}
  \includegraphics[height=0.8\columnwidth]{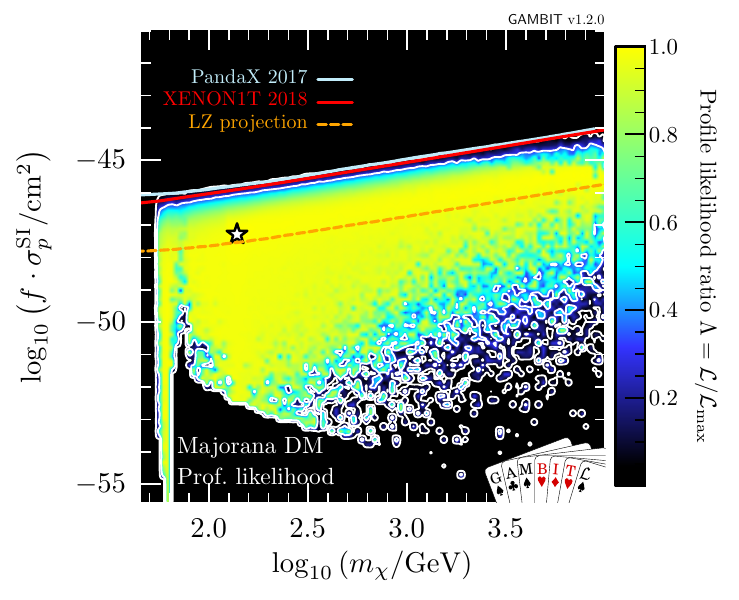}
  \includegraphics[height=0.8\columnwidth]{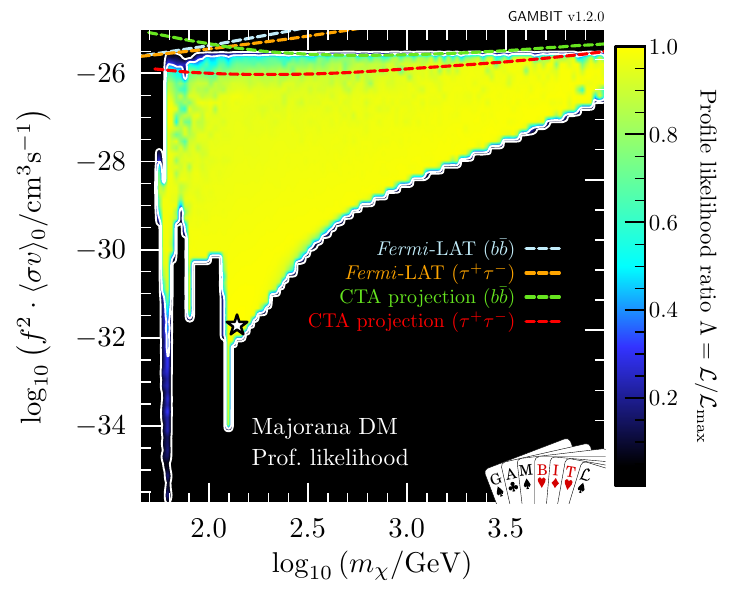}
  \caption{Same as Fig.~\ref{fig::vector_rescaled_obs} but for Majorana fermion DM.  Again, $f\equiv \Omega_{\chi}/\Omega_{\mathrm{DM}}$. For illustration, as there is a $q^2$-suppression in the spin-independent cross-section (see Eq.~\ref{eq:fermion_SI_xsec}), we show $\sigma_\mathrm{SI}$ computed at a reference momentum exchange of $q=50$\,MeV.}
  \label{fig::majorana_rescaled_obs}
\end{figure}

In the resonance region, we see the same triangular region as in the vector DM case: bounded from below by the relic density, and from the upper-left by the Higgs invisible width. However, in contrast to the vector DM case where direct detection limits squeeze the allowed region from the upper right, the addition of the mixing angle $\xi$ as a free parameter allows for the fermionic DM models to escape these constraints. As the pseudoscalar coupling is increased and the scalar coupling is correspondingly decreased, the SI cross-section becomes steadily more $q^2$-suppressed, as seen in Eq.~\eqref{eq:fermion_SI_xsec}. Noting that, the neck region at $m_\chi = m_h/2$ is less well-defined than in the vector and scalar DM cases above the triangle region. Notably however, as the SI cross-section becomes steadily more $q^2$-suppressed, the annihilation cross-section becomes \textit{less} $p$-wave suppressed (Eq.\ \ref{eq:prefactor}), and indirect detection comes to dominate the constraint at the edge of the allowed parameter space just above the resonance.

In the low-mass resonance region, virtually all values of the mixing angle are permitted, seen clearly in the left panel of Fig.~\ref{fig::majorana_mixing}, as even purely scalar couplings are not sufficient for direct detection to probe the remaining parameter space. The right panel also shows this in the lower `bulb': couplings between $10^{-3}\,\rm{GeV}^{-1}$ and $10^{-5}\,\rm{GeV}^{-1}$ are only permitted in the resonance region, without any constraint on the mixing angle.

In the high-mass region, we see that unlike the vector DM case, a wide range of WIMP masses between $100\,\text{GeV}$ and $10\,\text{TeV}$ are acceptable, with degenerate maximum likelihood. This is again due to the $q^2$-suppression of the direct detection constraints when considering all possible values of $\xi$. The large triangular high-mass region is constrained by the EFT validity constraint from above (highlighted in dark grey) and the relic density constraint from below.

\begin{figure*}[tbp]
  \centering
  \includegraphics[height=0.8\columnwidth]{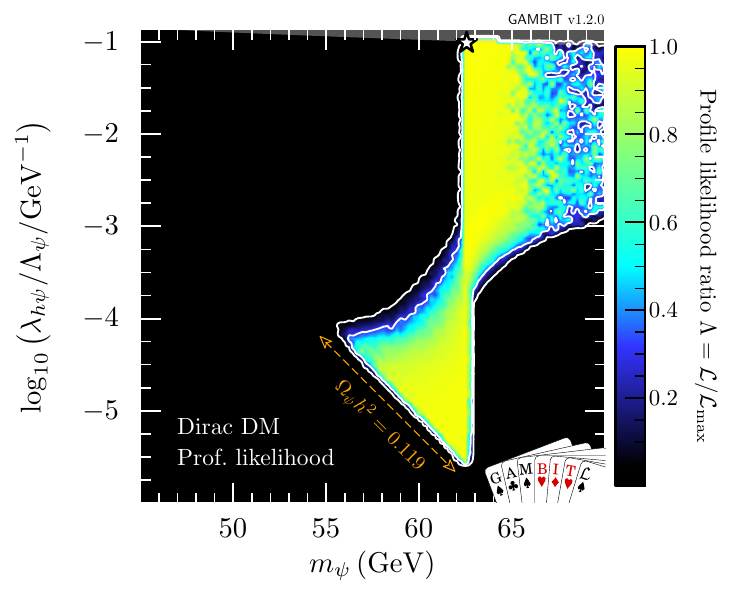}
  \includegraphics[height=0.8\columnwidth]{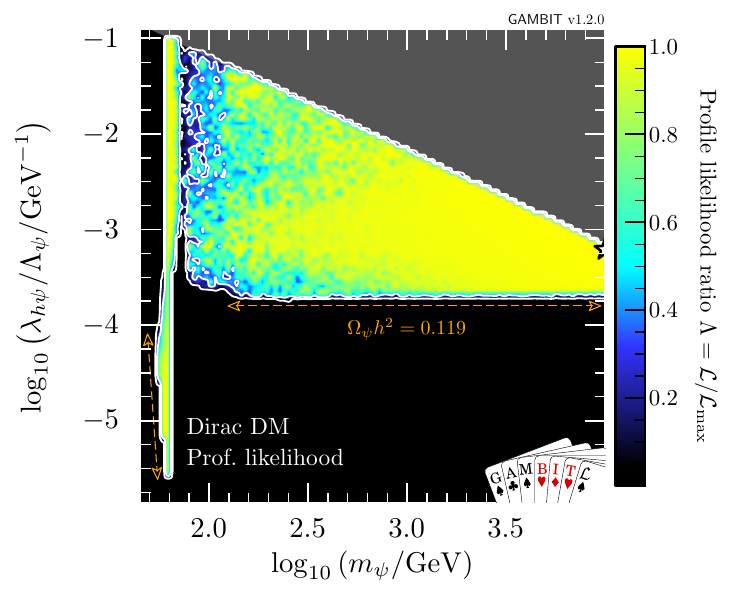}
  \vspace{-2mm}
  \caption{Profile likelihood in the $(m_{\psi},\lambda_{h\psi}/\Lambda_\psi)$ plane for Dirac fermion DM. Contour lines show the $1$ and $2\sigma$ confidence regions. The left panel gives an enhanced view of the resonance region around $m_{\psi} \sim m_h/2$. The right panel shows the full parameter space explored in our fits. The greyed out region shows where our approximate bound on the validity of the EFT is violated, the white stars show the best-fit point for each mass region, and the edges of the preferred parameter space along which the model reproduces the entire observed relic density are indicated with orange annotations.}
  \label{fig::dirac_profile}
\end{figure*}

\begin{table*}[tbp]
\center{
  \begin{tabular}{l l l l l l l}
  Model            & Relic density condition                   & $\lambda_{hX}$                       & $m_X$\,(GeV)       & $\xi$\, (rad) & $\Omega_Xh^2$ & $\Delta\ln\mathcal{L}$ \\
  \toprule
  Vector           & $\Omega_V h^2 \lesssim \Omega_{DM}h^2$    & $4.9\times 10^{-4}$                  & $62.46$            &  ---          & $9.343\times 10^{-2}$ & 0.322 \\
                   & $\Omega_V h^2 \sim \Omega_{DM}h^2$        & $4.5\times 10^{-4}$                  & $62.46$            &  ---          & $1.128\times 10^{-1}$ & 0.428 \\ \hline
  Majorana         & $\Omega_\chi h^2 \lesssim \Omega_{DM}h^2$ & $4.5\times 10^{-2}\,\text{GeV}^{-1}$ & $138.4$            &  1.96         & $6.588\times 10^{-8}$ & 0.308 \\
                   & $\Omega_\chi h^2 \sim \Omega_{DM}h^2$     & $6.3\times 10^{-6}\,\text{GeV}^{-1}$ & $61.03$            &  1.41         & $1.128\times 10^{-1}$ & 0.439 \\ \hline
  Dirac            & $\Omega_\psi h^2 \lesssim \Omega_{DM}h^2$ & $6.3\times 10^{-4}\,\text{GeV}^{-1}$ & $9.950\times 10^3$ &  2.06         & $3.813\times 10^{-2}$ & 0.307 \\
                   & $\Omega_\psi h^2 \sim \Omega_{DM}h^2$     & $3.6\times 10^{-4}\,\text{GeV}^{-1}$ & $9.895\times 10^3$ &  2.07         & $1.155\times 10^{-1}$ & 0.553 \\
  \bottomrule
  \end{tabular}
  \caption{Details of the best-fit parameter points for vector, Majorana and Dirac DM Higgs portal models, both with and without the requirement that the predicted relic density is within $1\sigma$ of the \emph{Planck} observed value. Here, $X \in \{V, \chi, \psi\}$ and the dimensionful nature of the coupling is implied for the fermion cases. We do not include the values of nuisance parameters, as they do not differ significantly from the central values of their likelihoods.}
  \label{tab::bestfit_params}
}
\end{table*}

In Fig.~\ref{fig::majorana_rescaled_obs}, we show the relic density (top) and scaled cross-sections for direct (centre) and indirect detection (bottom). For plotting purposes, we compute $\sigma_{\rm{SI}}$ at a reference momentum exchange of $q=50$\,MeV, typical of direct detection experiments. Substantial fractions of allowed parameter space lie close to current limits, but unsurprisingly, large portions of the parameter space will not be probed by future direct detection experiments, due to the momentum suppression. This is also true for indirect detection, where cross-sections are velocity suppressed.  However, given that the two suppressions have opposite dependences on the mixing parameter, the two probes will be able to compensate for each others' weaknesses to a certain extent.

Table~\ref{tab::bestfit} shows a breakdown of the contributions to the likelihood at the best-fit point, which lies in the high mass region at $m_\chi = 138.4$\,GeV, $\lambda_{h\chi}/\Lambda_\chi=4.5\times10^{-2}$\,{GeV}$^{-1}$ and $\xi = 1.96$\,rad (Table \ref{tab::bestfit_params}). When we demand that $\chi$ saturates the observed DM relic abundance, the best-fit point shifts to the lower end of the resonance region at $m_\chi = 61.03$\,GeV, $\lambda_{h\chi}/\Lambda_\chi=6.3\times10^{-6}$\,GeV$^{-1}$ and $\xi = 1.41$\,rad.

\subsubsection{Dirac fermion model}

The results from our low- and high-mass scans of the Dirac fermion model are very similar to those for the Majorana model. We therefore only show results in the $(m_\psi, \lambda_{h\psi}/\Lambda_\psi)$ plane in Fig.~\ref{fig::dirac_profile}.

In Table~\ref{tab::bestfit}, we show a breakdown of the contributions to the likelihood at the best-fit point.  This point lies towards the upper end of the high mass region, where $\lambda_{h\psi}/\Lambda_\psi=6.3\times10^{-4}$\,GeV$^{-1}$, $m_\psi = 9.95$\,TeV and $\xi = 2.06$\,rad. If $\psi$ makes up all of the DM, the best-fit point shifts slightly to the bottom of the high mass triangle at $\lambda_{h\psi}/\Lambda_\psi=3.6\times10^{-4}$\,GeV$^{-1}$, $m_\psi = 9.9$\,TeV and $\xi = 2.07$\,rad.  We compare the locations of these best-fit points to those from the vector and Majorana models in Table \ref{tab::bestfit_params}.

\begin{figure*}[tbp]
  \centering
  \includegraphics[height=0.8\columnwidth]{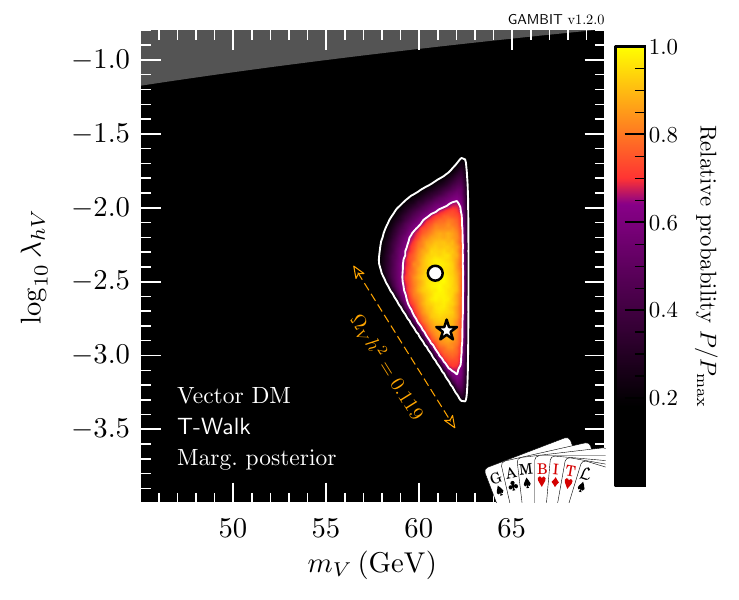}
  \includegraphics[height=0.8\columnwidth]{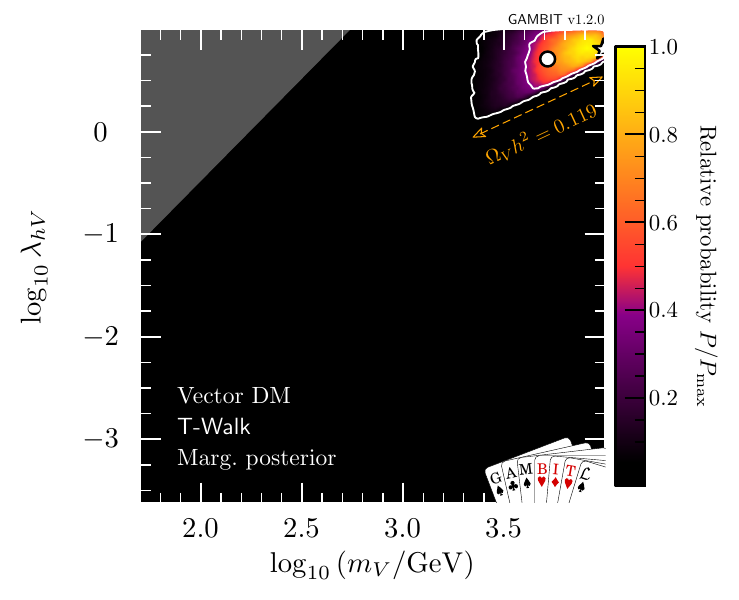}
  \vspace{-2mm}
  \caption{Marginalised posterior distributions in the $(m_V,\lambda_{hV})$ plane for vector DM. Contour lines show the $1$ and $2\sigma$ credible regions. The left panel gives the result of a scan restricted to the resonance region around $m_V \sim m_h/2$. The right panel shows a full-range parameter scan.   The low-mass mode is sufficiently disfavoured in the full-range scan that it does not appear in the righthand panel. The greyed out region shows points that do not satisfy Eq.~\eqref{eq:pert_unitarity}. The posterior mean is shown by a white circle, while the maximum likelihood point is shown as a white star.  The edges of the preferred parameter space along which the model reproduces the entire observed relic density are indicated with orange annotations.}
  \label{fig::vector_TWalk}
\end{figure*}

\subsubsection{Goodness of fit}

In Table~\ref{tab::bestfit}, we show the contribution to the log-likelihood for the best-fit points of the vector, Majorana and Dirac DM models. By equating $\Delta\ln\mathcal{L}$ to half the ``likelihood $\chi^2$'' of Baker \& Cousins \cite{Baker:1983tu}, we can compute an approximate $p$-value for each best-fit point against a null hypothesis. We take this null to be the `ideal' case, which we define as the background-only contribution in the case of exclusions, and the observed value in the case of detections.

For the vector DM model, using either one or two effective degrees of freedom, we find a $p$-value between roughly $0.4$ and $0.7$. Requiring the relic density of $V_\mu$ to be within $1\sigma$ of the \emph{Planck} value, the $p$-value becomes $p\approx 0.35$--$0.65$. For both the Majorana and Dirac fermion models, we also find $p \approx 0.4$--$0.7$, falling to $0.35$--$0.65$ with the relic density requirement. All of these are completely acceptable $p$-values.

\begin{figure*}[tbp]
  \centering
  \includegraphics[height=0.8\columnwidth]{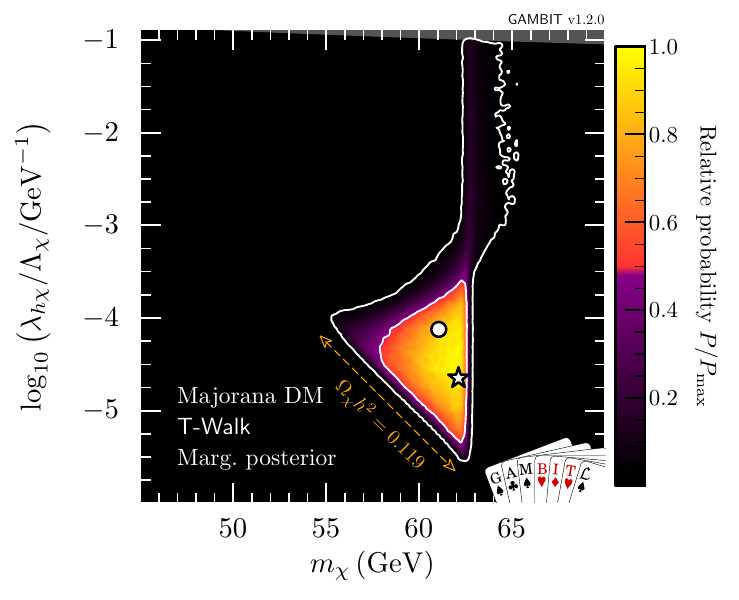}
  \includegraphics[height=0.8\columnwidth]{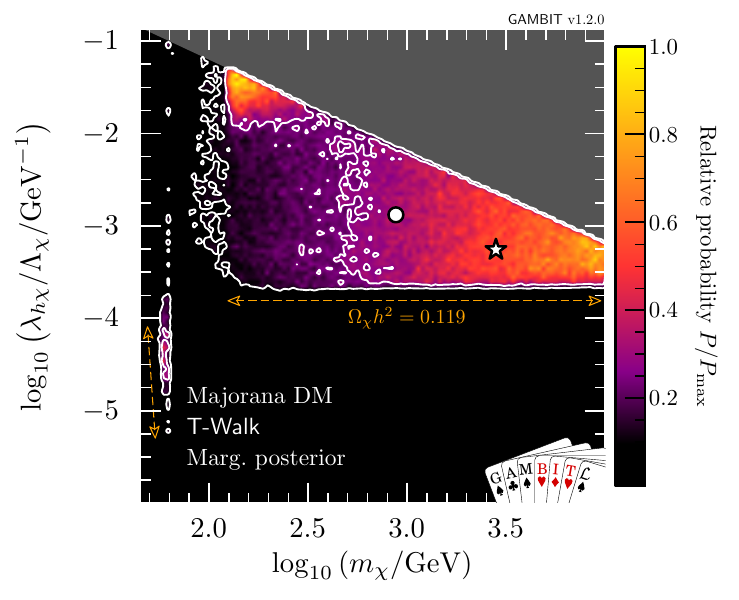}
  \vspace{-2mm}
  \caption{Marginalised posterior distributions in the $(m_\chi,\lambda_{h\chi}/\Lambda_\chi)$ plane for Majorana fermion DM. Contour lines show the $1$ and $2\sigma$ credible regions. The left panel gives the result of a scan restricted to the resonance region around $m_\chi \sim m_h/2$. The right panel shows a full-range parameter scan.  The greyed out region shows where our approximate bound on the validity of the EFT is violated. The posterior mean is shown by a white circle, while the maximum likelihood point is shown as a white star. The edges of the preferred parameter space along which the model reproduces the entire observed relic density are indicated with orange annotations.}
  \label{fig::majorana_TWalk}
\end{figure*}

\begin{figure*}[tbp]
 \centering
 \includegraphics[height=0.8\columnwidth]{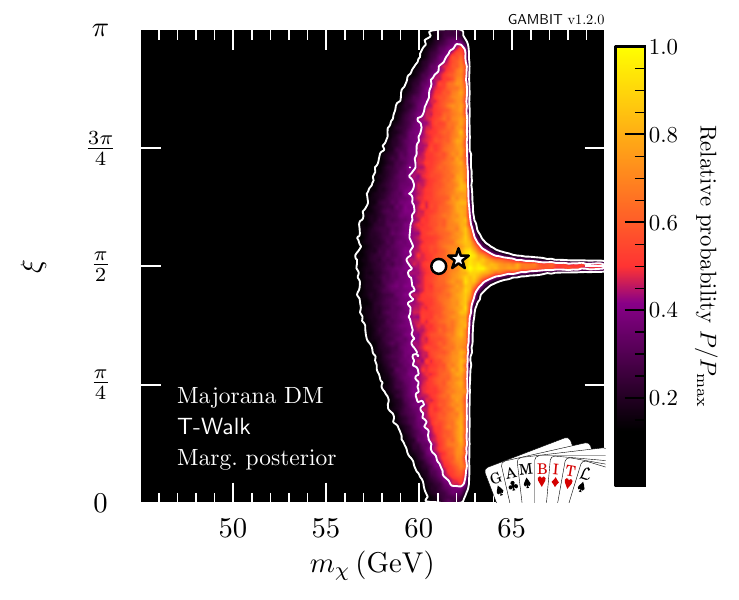}
 \includegraphics[height=0.8\columnwidth]{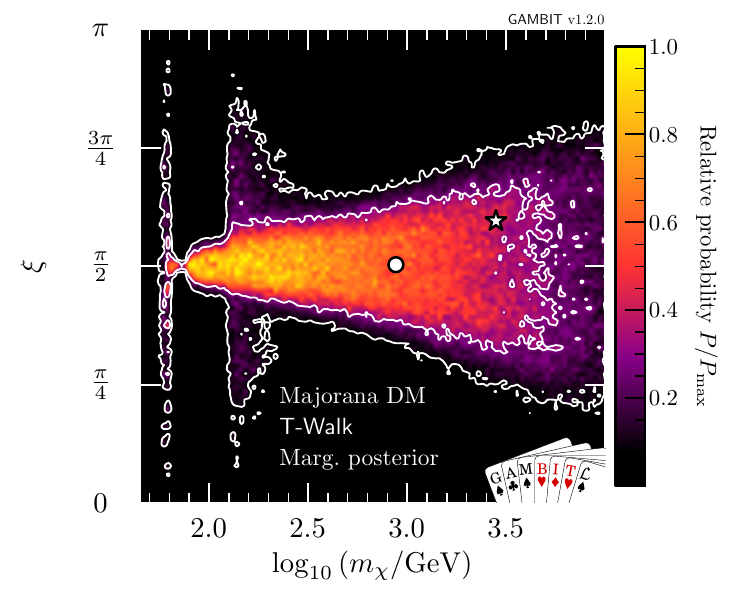}
  \vspace{-2mm}
  \caption{Marginalised posterior distributions in the $(m_\chi,\xi)$ plane for Majorana fermion DM. Contour lines show the $1$ and $2\sigma$ credible regions. The left panel gives the result of a scan restricted to the resonance region around $m_\chi \sim m_h/2$. The right panel shows a full-range parameter scan. The posterior mean is shown by a white circle, while the maximum likelihood point is shown as a white star.}
 \label{fig::majorana_mixing_angle}
\end{figure*}

\begin{figure*}[tbp]
 \centering
 \includegraphics[height=0.8\columnwidth]{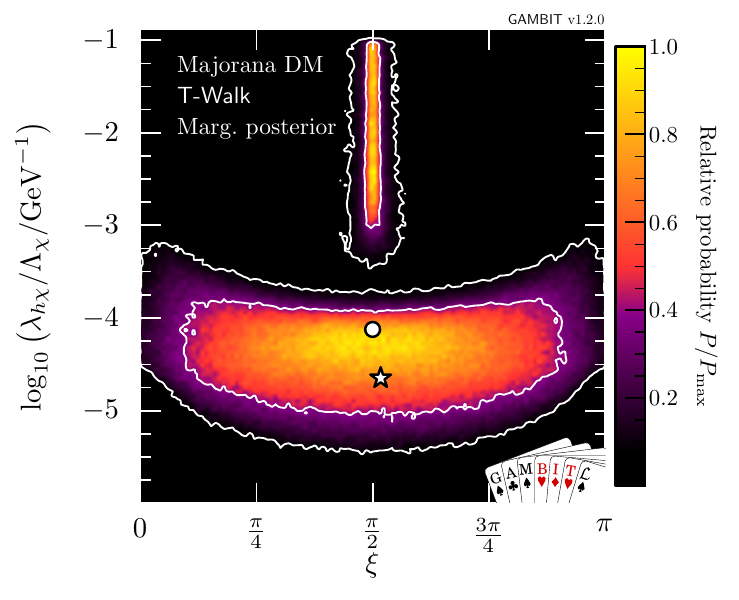}
 \includegraphics[height=0.8\columnwidth]{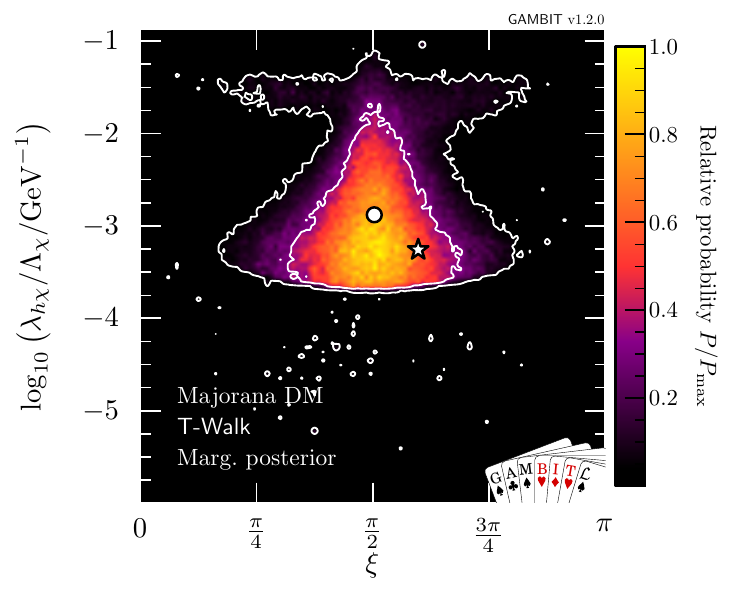}
  \vspace{-2mm}
  \caption{Marginalised posterior distributions in the $(\lambda_{h\chi}/\Lambda_\chi,\xi)$ plane for Majorana fermion DM. Contour lines show the $1$ and $2\sigma$ credible regions. The left panel gives the result of a scan restricted to the resonance region around $m_\chi \sim m_h/2$. The right panel shows a full-range parameter scan.  The posterior mean is shown by a white circle, while the maximum likelihood point is shown as a white star.}
 \label{fig::majorana_bulb_post}
\end{figure*}

\subsection{Marginal posteriors}
\label{posteriors}

The marginal posterior automatically penalises fine-tuning, as upon
integration of the posterior, regions with a limited `volume of
support' over the parameters that were integrated over are
suppressed.\footnote{By `volume of support', we mean the regions of
the parameter space that have a non-negligible likelihood times prior
density.} As usual, the marginal posteriors depend upon the choice of
priors for the free model parameters, which are summarised in
Tables~\ref{tab:vector_params} and \ref{tab:fermi_params}.  We choose
flat priors where parameters are strongly restricted to a particular
scale, such as the mixing parameter and the DM mass in scans
restricted to the low-mass region.  For other parameters, in order to
avoid favouring a particular scale we employ logarithmic
priors. Note that in this treatment for the fermionic DM
models we have not chosen priors that favour the CP-conserving case.
We instead present posteriors for this well motivated case separately,
and later in section \ref{sec:evidence} we perform a Bayesian model
comparison between a CP-conserving fermionic DM model and the full
model considered here.

\subsubsection{Vector model}

To obtain the marginal posterior distributions, we perform separate \twalk scans for the low and high mass regimes, shown in Fig.~\ref{fig::vector_TWalk}. Within each region we plot the relative posterior probability across the parameter ranges of interest.

In the left panel of Fig.~\ref{fig::vector_TWalk}, the scan of the resonance region shows that the neck region is disfavoured after marginalising over the nuisance parameters, particularly $m_h$, which sets the width of the neck. This dilutes the allowed region due to volume effects.

In the full-mass-range scan, the fine-tuned nature of the resonance region is clearly evident. Although the best-fit point in the profile likelihood lies in the resonance region, the posterior mass is so small in the entire resonance region that it drops out of the global $2\sigma$ credible interval.

\subsubsection{Majorana fermion model}

As already seen in the profile likelihoods, for the case of Majorana fermion DM, the presence of the mixing parameter $\xi$ leads to a substantial increase in the preferred parameter region (see Fig.~\ref{fig::majorana_TWalk}). In the resonance region (left panel), there is now a thin neck-like region at $m_\chi \approx m_h / 2$. This neck region is the same one seen in both the scalar and vector profile likelihoods, but falls within the $2\sigma$ credible region of the Majorana posterior, as the admittance of $\xi$ reduces direct detection constraints (Eq.\ \ref{eq:fermion_SI_xsec}), softening the penalisation from integrating over nuisance parameters. When we compute the posterior over the full mass range, we once again find the resonance region to be somewhat disfavoured, but now there are large parameter regions with high posterior probabilities for $m_\chi > m_h$.

Nevertheless, direct detection does have a significant impact on the high-mass region, in spite of the mixing parameter $\xi$. While the $2 \sigma$ contour is roughly triangular, the points with highest posterior probability (i.e.\ within the $1 \sigma$ contours) are split into two smaller triangles. The approximately rectangular region that separates these two triangular regions is disfavoured by the combination of volume effects and direct detection, which requires $\xi$ to be tuned relatively close to $\pi/2$.

To better understand the role of tuning in $\xi$ in the process of marginalisation, we show the marginalised posterior in the $(m_\chi, \xi)$ and $(\xi,\lambda_{h\chi}/\Lambda_\chi)$ planes in Figs.~\ref{fig::majorana_mixing_angle} and \ref{fig::majorana_bulb_post}, respectively. Fig.~\ref{fig::majorana_mixing_angle} provides a clear understanding of the differences between the marginalised posteriors in Fig.~\ref{fig::majorana_TWalk} and the profile likelihood in Fig.~\ref{fig::majorana_profile}. In the resonance region (left panel), the neck region is less prominent in the marginalised posterior because direct detection limits become very constraining as soon as $m_\chi > m_h/2$ and the mixing parameter is forced to be very close to $\pi/2$. In the full-range scan (right panel) we see the annihilation channel $\ovr{\chi} \chi \to h h$ open up, thus allowing a greater range of values for $\xi$, leading to an enhancement in the marginalised posterior probability. This clearly corresponds to the $1 \sigma$ triangular region in the mass-coupling plane at $m_\chi \approx m_h$, in the right hand panel of Fig.~\ref{fig::majorana_TWalk}.

In the left panel of Fig.~\ref{fig::majorana_bulb_post}, which focuses on the resonance region, we see two separate solutions for the mixing angle and coupling: the larger island at lower coupling corresponds to the triangular region at $m_\chi < m_h/2$, permitting all values of $\xi$, and the thinner solution at larger couplings reflects the solution at $m_\chi > m_h/2$, where the scalar coupling between the Higgs and the Majorana DM needs to be sufficiently small (i.e. $\xi \sim \pi/2$) to evade direct detection limits. The two regions appear disconnected because the intermediate parameter points require so much tuning that they fall outside of the $2\sigma$ credible regions upon marginalisation. Considering the full mass range (see the right panel in Fig.~\ref{fig::majorana_bulb_post}), we find that the lower `bulb' seen in the profile likelihood in Fig.~\ref{fig::majorana_mixing} is hardly visible in the marginalised posterior when integrating over the nuisance parameters, due to a lower posterior volume in the resonance region.

\begin{figure}[tbp]
 \centering
 \includegraphics[height=0.8\columnwidth]{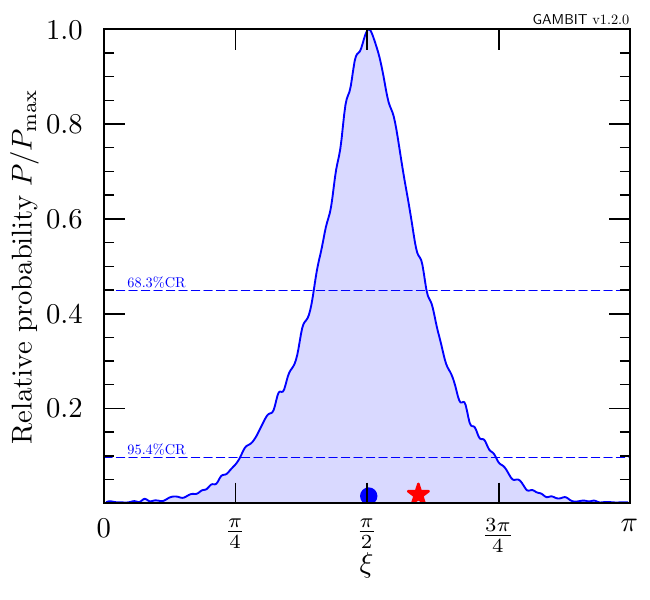}
 \caption{Marginalised posterior distribution for the mixing angle $\xi$ for Majorana fermion DM in the full-mass-range scan.
 The posterior mean is shown by a blue circle, while the maximum likelihood point is shown as a red star.}
 \label{fig::majorana_xi_1D_posterior}
\end{figure}

We can condense the information from Figs.~\ref{fig::majorana_mixing_angle} and~\ref{fig::majorana_bulb_post} further by marginalising over all parameters except for $\xi$, thus obtaining a 1D posterior probability. The result is shown in Fig.~\ref{fig::majorana_xi_1D_posterior}, where the preference for $\xi\approx\pi/2$ becomes clear. In other words, for the case of Majorana fermion DM, there is a strong preference for permitting an increased admixture of pseudoscalar-type couplings to suppress the constraints from direct detection and the relic density, due to a momentum and velocity suppressed cross-section respectively.

\begin{figure*}[tbp]
 \centering
 \includegraphics[height=0.8\columnwidth]{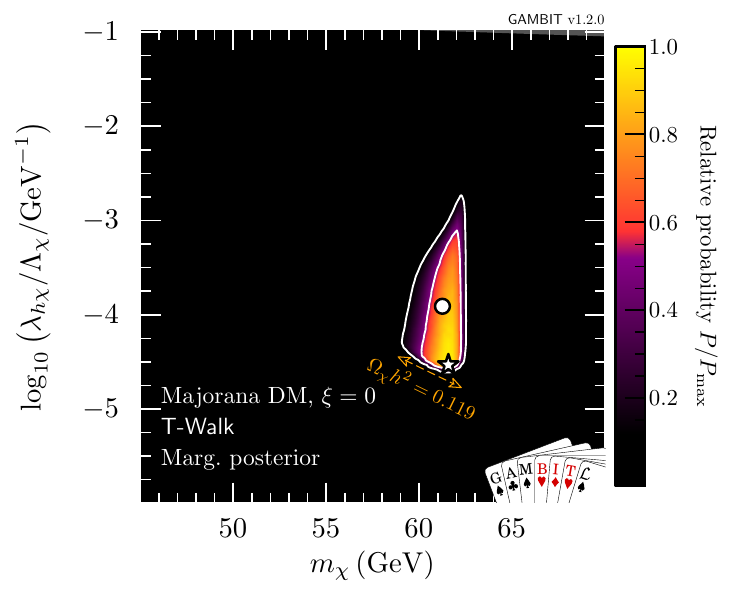}
 \includegraphics[height=0.8\columnwidth]{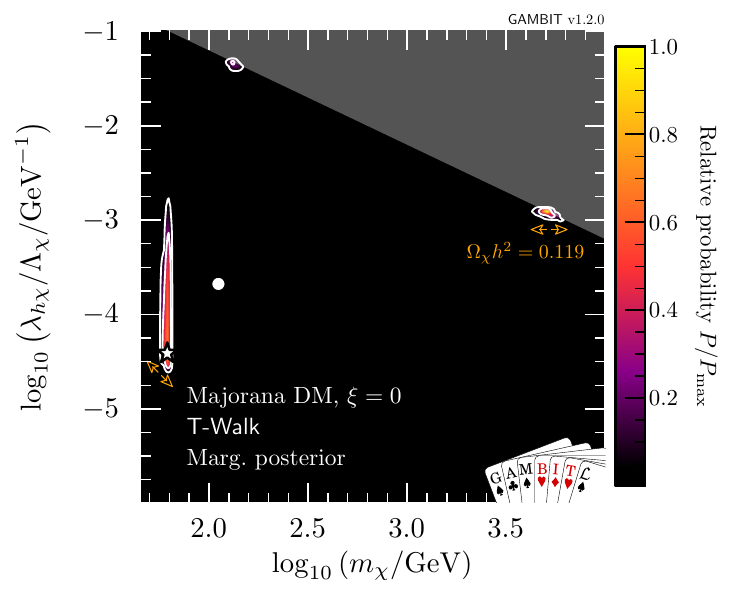}
  \vspace{-2mm}
  \caption{Marginalised posterior distributions for Majorana fermion DM with fixed $\xi=0$. Contour lines show the $1$ and $2\sigma$ credible regions. The left panel gives the result of a scan restricted to the resonance region around $m_\chi \sim m_h/2$. The right panel shows a full-range parameter scan. The posterior mean is shown by a white circle, while the maximum likelihood point is shown as a white star. The edges of the preferred parameter space along which the model reproduces the entire observed relic density are indicated with orange annotations.}
 \label{fig::majorana_fixed_xi}
\end{figure*}

For comparison, we consider the CP-conserving case with fixed $\xi = 0$ in Fig.~\ref{fig::majorana_fixed_xi}. As expected from the discussion above, we find that the permitted parameter space shrinks vastly with respect to the case where the mixing parameter is allowed to vary (see Fig.~\ref{fig::majorana_TWalk}). In the resonance region (left panel), we see that direct detection, the invisible Higgs width and relic density impose strong constraints from the left, upper-left and below, respectively. No neck region exists because the direct detection constraints are too strong, overlapping with constraints on the invisible width of the Higgs boson. In the full-range scan (right panel), we find that the only surviving parameter space is split into the resonance region, and two small islands, at $m_\chi \sim m_h$ and $m_\chi \sim 5$\,TeV. These islands are constrained by direct detection and the EFT validity requirement. Both will be ruled out by the next generation of direct detection experiments, if no DM signal is observed.

Our analysis of the Dirac fermion model parameter space is identical to the Majorana fermion one, whether $\xi$ is fixed or left as a free parameter, so to avoid repetition we omit those results.

It should be clear from the comparison between Figs.~\ref{fig::majorana_TWalk} and ~\ref{fig::majorana_fixed_xi} that the CP-conserving case $(\xi = 0)$ is strongly disfavoured relative to the case where $\xi$ is allowed to vary. We will make this qualitative observation more precise in the following section.

\section{Bayesian model comparison} \label{sec:evidence}

\subsection{Background}

To be able to comment on the relative plausibility of the different Higgs portal models, we must also perform a quantitative model comparison.  To do this, we compute Bayes factors for pairs of models, say $\mathcal{M}_1$ and $\mathcal{M}_2$ as \cite{Jeffreys:1939xee,10.2307/2291091,10.2307/4356165}
\begin{equation}
B \equiv \frac{\mathcal{Z}(\mathcal{M}_1)}{\mathcal{Z}(\mathcal{M}_2)} \, ,
\label{eq:bayes_factor}
\end{equation}
where $\mathcal{Z}(\mathcal{M})$ is the evidence of a model $\mathcal{M}$.  This is the integral of the likelihood of the observed data $\mathcal{L}(D|\theta)$ over the possible parameter values $\theta$ in that model, weighted by the prior on the parameters $P(\theta)$,
\begin{equation}
\mathcal{Z}(\mathcal{M}) \equiv \int \mathcal{L}(D|\theta) P(\theta) \, d\theta \, .
\label{eq:evidence}
\end{equation}
We perform this integration using \multinest \cite{Feroz:2007kg,Feroz:2008xx}, which is designed to calculate the Bayesian evidence. The final odds ratio (of the probability that $\mathcal{M}_1$ is correct to the probability that $\mathcal{M}_2$ is correct) is the product of the Bayes factor and the ratio of any prior beliefs $P(\mathcal{M}_1)/P(\mathcal{M}_2)$ that we might have in these models,
\begin{equation}
\frac{P(\mathcal{M}_1|D)}{P(\mathcal{M}_2|D)} = B \frac{P(\mathcal{M}_1)}{P(\mathcal{M}_2)} \, .
\end{equation}
In our analysis, we take the prior probability of every model to be equal such that the factor,
\begin{equation}
P(\mathcal{M}_1)/P(\mathcal{M}_2) = 1
\label{Eq:equal_priors}
\end{equation}
for all pairs of models. Thus, the odds ratio is simply given by the Bayes factor.  Note that even when computing the evidence for or against CP violation in the fermionic model below, we do \textit{not} apply any further prior in favour of CP conservation.  The volume integrals involved in the Bayes factor automatically implement the concept of naturalness via Occam's razor, penalising models with more free parameters if they do not fit the observed data any better than models with less parameters.

From Eq.~\eqref{eq:evidence}, we can see that the evidence of a model depends on the prior choices for its parameters. This prior on the model parameters (along with the priors on the models themselves) makes the results of Bayesian model comparison inherently prior-dependent.  However, the influence of common parameters treated with identical priors in both models approximately cancels when taking the ratio of evidences, as in Eq.~\eqref{eq:bayes_factor}. The overall prior dependence of the Bayes factor can thus be minimised by minimising the number of non-shared parameters between the models being compared. The best case is where one model is nested inside the other, and corresponds simply to a specific choice for one of the degrees of freedom in the larger model.  In this case, the leading prior dependence is the one coming from the chosen prior on the non-shared degree of freedom. Thus, we first investigate the question of CP violation in the Higgs portal, which we can address in this manner, before going on to the more prior-dependent comparison of the broader models.

\subsection{CP violation in the Higgs portal}\label{sec:bayes_factor_CP}

We perform Bayesian model comparison for the fermionic Higgs portal DM, and nested variants of it, by comparing the CP-conserving case ($\xi =0$) to the model where the CP phase of the portal coupling is allowed to vary freely. Due to the similarity of the likelihood for the Dirac and Majorana fermion models, we do this for the Majorana fermion model only.  We carry out this exercise for two different parametrisations of the model, corresponding to two different priors on the larger parameter space in which the CP-conserving scenario is nested:
\begin{enumerate}
\item Assuming the parametrisation that we have discussed thus far for the Majorana model, taking a uniform prior for $\xi$ and a logarithmic prior for $\lambda_{h\chi}/\Lambda_\chi$.  This corresponds to the assumption that some single mechanism uniquely determines the magnitude and phase of both couplings.
\item Assuming that the scalar and pseudoscalar couplings originate from distinct physical mechanisms at unrelated scales, such that they can be described by independent logarithmic priors. The post-EWSB Lagrangian in this parametrisation contains the terms
  \begin{equation}
    \lagr_\chi \supset - \frac12\left( \frac{g_\mathrm{s}}{\Lambda_\mathrm{s}} \ovr{\chi} \chi + \frac{g_\mathrm{p}}{\Lambda_\mathrm{p}} \ovr{\chi} i\gamma_5 \chi \right) \left(\vo h + \frac{1}{2} h^2 \right) \, .
  \end{equation}
  In this case, the parameters $\xi$ and $\lambda_{h\chi}/\Lambda_\chi$ from the first parametrisation are replaced by $g_\mathrm{s}/\Lambda_\mathrm{s}$ and $g_\mathrm{p}/\Lambda_\mathrm{p}$.  In this parametrisation, the Bayes factor may be sensitive to the range of the prior for the couplings, as the normalisation factor does not cancel when computing the Bayes factor for the CP-conserving scenario.  We choose $-6 \le \log_{10} ({g/\Lambda}) \le 0$ for the couplings when computing the Bayes factors in this parametrisation, in line with the prior that we adopt for $\lambda_{h\chi}/\Lambda_\chi$ in parametrisation 1.
\end{enumerate}
The CP-conserving model is nested within both of these models, as $\xi=0$ in the first, and as $g_\mathrm{p}/\Lambda_\mathrm{p}=0$ in the second (although the exact limit of $\xi=0$ is not contained within our chosen prior for the second parameterisation, seeing as we choose a logarithmic prior on $g_\mathrm{p}$). As stated in Eq.~\ref{Eq:equal_priors}, the ratio of the prior probabilities for these models is taken to be $1$ here, and is not related to priors of parameters discussed above. We are comparing two separate models: one with a pure CP-even coupling between the DM fermion and the Higgs and another model where there is also a pseudoscalar coupling, which {\it a priori} is very unlikely to be zero.

In Table~\ref{tab::bfa}, we give the odds ratios against the CP-conserving case in each of these parametrisations. The value given in the final column of this table is the ratio of the evidence for the CP-violating model to the CP-conserving case.  Depending on the choice of parametrisation, we see that there is between 140:1 and 70:1 odds against the CP-conserving version of the Majorana Higgs portal model. The similarity in order of magnitude\footnote{Odds ratios are best conceived of in a logarithmic sense, so a factor of 2 difference is of negligible importance.} between these two results is expected, as it reflects the relatively mild prior-dependence of the Bayes factor when performing an analysis of nested models that differ by only a single parameter. Given the similarity of the likelihood functions in the Majorana and Dirac fermion models, the odds against the pure CP-conserving version of the Dirac fermion Higgs portal model can also be expected to be very similar.

The odds ratio tells us the relative plausibility of one model relative to the other. According to the standard scale frequently used for interpreting Bayesian odds ratios (the Jeffreys scale; \cite{Jeffreys:1939xee,10.2307/2291091}), this constitutes strong evidence against pure CP-even coupling in fermionic Higgs portal models. The preference for a CP-violating coupling can also be seen in Fig.~\ref{fig::majorana_xi_1D_posterior}, where there is a clear preference for $\xi = \pi/2$, whereas the CP-even coupling falls outside of the $2\sigma$ credible region.

\begin{table}[tbp]
  \centering
  \begin{tabular}{ccc}
    \hline
    Model & Comparison model and priors & Odds \\ \hline
    $\xi=0$ & $m_\chi$: log \quad $\lambda_{h\chi}/\Lambda_\chi$: log \quad $\xi$: flat & 70:1 \\[1mm]
    $g_\mathrm{p}/\Lambda_\mathrm{p} = 0$ & $m_\chi$: log \quad $g_\mathrm{s}/\Lambda_\mathrm{s}$: log \quad $g_\mathrm{p}/\Lambda_\mathrm{p}$: log & 140:1 \\
    \hline
  \end{tabular}
  \caption{Odds ratios for CP violation for the singlet Majorana fermion Higgs portal model. Here the odds ratios are those against a pure CP-even Higgs portal coupling, as compared to two different parametrisations (and thus priors) of the model in which the CP nature of the Higgs portal can vary freely. \label{tab::bfa}}
\end{table}

\subsection{Scalar, Vector, Majorana or Dirac?}\label{sec:bayes_factor_spin}

We also carry out model comparison between the different Higgs portal models: scalar, vector, Majorana and Dirac. As these models are not nested, they each have unique parameters.  This means that there is no \textit{a priori} relationship between their respective parameters that would allow the definition of equivalent priors on, e.g., masses or couplings in two different models.  The prior dependence of the Bayes factor is therefore unsuppressed by any approximate cancellations when taking the ratio of evidences in Eq.~\eqref{eq:bayes_factor}. We caution that the resulting conclusions are consequently less robust than for the nested Majorana models. For this exercise, we update the fit to the scalar model from Ref.\ \cite{SSDM} to incorporate the likelihood function and nuisances that we use in the current paper.

We find that the scalar Higgs portal model has the largest evidence value in our scans, but is comparable to the fermion DM models. In Table~\ref{tab::bfb}, we give the odds ratios against each of the Higgs portal models, relative to the scalar model. The data have no preference between scalar and either form of fermionic Higgs portal model, with odds ratios of 1:1. The vector DM model is disfavoured with a ratio of 6:1 compared to the scalar and fermion models; this constitutes `positive' evidence against the vector DM model according to the Jeffreys scale, though the preference is only rather mild.  Overall, there is no strong preference for Higgs portal DM to transform as a scalar, vector or fermion under the Lorentz group.

As we find no strong preference between the different Higgs portal DM models using logarithmic priors, we omit a dedicated prior sensitivity analysis. If different assumptions on priors were to yield a stronger preference for any of the models under consideration, the only conclusion would be that such a preference is not robust to changes in the prior. The situation is hence different from the one in Sec.~\ref{sec:bayes_factor_CP}, where we did find a strong preference against the CP-conserving model, which we showed to be largely independent of the assumed prior.

\begin{table}[tbp]
  \centering
  \begin{tabular}{ccc}
    \hline
    Model   & Parameters and priors & Odds \\ \hline
    $S$       & $m_S$: log \quad   $\lambda_{hS}$: log                               & 1:1 \\[1mm]
    $V_{\mu}$ & $m_V$: log  \quad $\lambda_{hV}$: log                 &               6:1 \\[1mm]
    $\chi$    & $m_\chi$: log  \quad $\lambda_{h\chi}/\Lambda_\chi$: log  \quad $\xi$: flat & 1:1 \\[1mm]
    $\psi$    & $m_\psi$: log  \quad $\lambda_{h\psi}/\Lambda_\psi$: log  \quad $\xi$: flat & 1:1 \\
    \hline
  \end{tabular}
  \caption{Odds ratios against each singlet Higgs portal DM model with $\mathbb{Z}_2$ symmetry, relative to the scalar model. \label{tab::bfb}}
\end{table}

%%%%%%%%%%%%%%%%%%%%%%%%%%%%%%%%%%%%%%%%%%%%%%%%%%%%%%%%%%%%%%%%%%%%%%%%%%%%%%%%%%%%%%%%%%%%%%%%%%%%%%%%%%%%%%%%%%%%%%%%
\section{Conclusions}\label{sec:conclusions}

In this study we have considered and compared simple extensions of the SM with fermionic and vector DM particles stabilised by a $\mathbb{Z}_2$ symmetry. These models are non-renormalisable, and the effective Higgs-portal coupling is the lowest-dimension operator connecting DM to SM particles. Scenarios of this type are constrained by the DM relic density predicted by the thermal freeze-out mechanism, invisible Higgs decays, and direct and indirect DM searches. Perturbative unitarity and validity of the corresponding EFT must also be considered.

We find that the vector, Majorana and Dirac models are all phenomenologically acceptable, regardless of whether or not the DM candidate saturates the observed DM abundance. In particular, the resonance region (where the DM particle mass is approximately half the SM Higgs mass) is consistent with all experimental constraints and challenging to probe even with projected future experiments. On the other hand, larger DM masses are typically tightly constrained by a combination of direct detection constraints, the relic density requirement and theoretical considerations such as perturbative unitarity. Our results show that with the next generation of direct detection experiments (e.g.,~LZ~\cite{LZ}), it will be possible to fully probe the high-mass region for both the vector and CP-conserving fermion DM model. Future indirect experiments such as CTA~\cite{Acharya:2017ttl} will be sensitive to parts of viable parameter space at large DM masses, but will have difficulty in probing the resonance region.

An interesting alternative is fermionic DM with a CP-violating Higgs portal coupling, for which the scattering rates in direct detection experiments are momentum-suppressed. By performing a Bayesian model comparison, we find that data strongly prefers the model with CP violation over the CP-conserving one, with odds of order 100:1 (over several priors). This illustrates how increasingly tight experimental constraints on weakly-interacting DM models are forcing us to abandon the simplest and most theoretically appealing models, in favour of more complex models.

We have also used Bayesian model comparison to determine the viability of the scalar Higgs portal model relative to the fermionic and vector DM models. We find a mild preference for scalar DM over vector DM, but no particular preference between the scalar and the fermionic model. This conclusion may however quickly change with more data. Stronger constraints on the Higgs invisible width will further constrain the resonance region and the combination of these constraints with future direct detection experiments may soon rule out the vector model.

Our study clearly demonstrates that, in the absence of positive signals, models of weakly-interacting DM particles will only remain viable if direct detection constraints can be systematically suppressed. This makes it increasingly interesting to study DM models with momentum-dependent scattering cross-sections. A systematic study of such theories will be left for future work. Conversely, Higgs portal models provide a natural framework for interpreting signals in the next generation of direct and indirect detection experiments. An advanced framework for such a reinterpretation using Fisher information will be implemented in future versions of \gambit.

%%%%%%%%%%%%%%%%%%%%%%%%%%%%%%%%%%%%%%%%%%%%%%%%%%%%%%%%%%%%%%%%%%%%%%%%%%%%%%%%%%%%%%%%%%%%%%%%%%%%%%%%%%%%%%%%%%%%%%%%
\begin{acknowledgements}
We thank Shyam Balaji, Archil Kobakhidze and Ross Young for helpful discussions, and Lucien Boland, Sean Crosby and Goncalo Borges from CoEPP Research Computing for providing computing assistance and allocating resources. We acknowledge PRACE for awarding us access to Marconi at CINECA, Italy. We are grateful to the UK Materials and Molecular Modelling Hub for computational resources, which is partially funded by EPSRC (EP/P020194/1).  This work was supported by STFC (1810964, ST/K00414X/1, ST/N000838/1, ST/P000762/1), the Swedish Research Council (contract 621-2014-5772), the Norwegian Research Council (FRIPRO project 230546/F20), NOTUR (Norway; NN9284K), DFG (Emmy Noether Grant No.\ KA 4662/1-1), NSERC, the \ Research Excellence Fund (CFREF), H2020 (ERC Starting Grant `NewAve' 638528, Marie Sk\l{}odowska-Curie Individual Fellowship `DarkGAMBIT' 752162), ARC (CE110001104 CoEPP, Future Fellowships FT140100244, FT160100274), the Australian Postgraduate Award and the Centre for the Subatomic Structure of Matter (CSSM).
\end{acknowledgements}

%%%%%%%%%%%%%%%%%%%%%%%%%%%%%%%%%%%%%%%%%%%%%%%%%%%%%%%%%%%%%%%%%%%%%%%%%%%%%%%%%%%%%%%%%%%%%%%%%%%%%%%%%%%%%%%%%%%%%%%%
\bibliography{R1.5}

%%%%%%%%%%%%%%%%%%%%%%%%%%%%%%%%%%%%%%%%%%%%%%%%%%%%%%%%%%%%%%%%%%%%%%%%%%%%%%%%%%%%%%%%%%%%%%%%%%%%%%%%%%%%%%%%%%%%%%%%
\appendix
\setcounter{table}{0}
\renewcommand\thetable{A\arabic{table}}

%%%%%%%%%%%%%%%%%%%%%%%%%%%%%%%%%%%%%%%%%%%%%%%%%%%%%%%%%%%%%%%%%%%%%%%%%%%%%%%%%%%%%%%%%%%%%%%%%%%%%%%%%%%%%%%%%%%%%%%%
\section{New features in DDCalc}\label{sec:DDCalc_feat}

In this appendix, we discuss the new features of \ddc, namely the treatment of general non-relativistic effective operators and the extended interface for implementing new analyses. For a more detailed illustration of the new features, we refer to the example programs in \path{DDCalc/examples/}, which are provided in both \texttt{C++} and \texttt{Fortran90}.\footnote{Note that \ddc no longer maintains a command line interface, so that the example files are in fact the only executables that are generated when compiling \ddcalc.} For an introduction into the basic structure of \ddcalc, we refer to Ref.~\cite{DarkBit}.

\subsection{Non-relativistic effective operators}

Up to second order in velocity and momentum transfer, elastic scattering of DM particles off nucleons via the exchange of a heavy mediator can be fully described by a set of 18 effective operators. These operators are conventionally denoted by $\mathcal{O}_1$, $\mathcal{O}_3$, \ldots, $\mathcal{O}_{15}$, $\mathcal{O}_{17}$, $\mathcal{O}_{18}$ (note that $\mathcal{O}_2$ and $\mathcal{O}_{16}$ are commonly omitted), as well as $q^2 \mathcal{O}_1$ and $q^2 \mathcal{O}_4$~\cite{Fitzpatrick:2012ix,Anand:2013yka,Dent:2015zpa}. Each of these operators can arise independently for scattering off protons and neutrons or, equivalently, for the isoscalar ($\tau = 0$) and the iso-vector ($\tau = 1$) current. As the interpretation of these operators also depends on the total spin $s_\chi$ of the DM particle, the interactions of DM are fully specified by a total of 37 parameters.

In order to consider a WIMP with general coupling structure, the user first initialises a generic WIMP object and then passes this object to specialised functions that define the coupling structure. For example, the following code initialises a WIMP with mass 50 GeV and spin $1/2$, and sets the isoscalar and iso-vector coefficients of the operator $\mathcal{O}_3$ to $0.1 \; \mathrm{GeV^{-2}}$ and $0.2 \; \mathrm{GeV^{-2}}$, respectively:\footnote{The normalisation of the non-relativistic operators corresponds to a DM particle that is not self-conjugate. Hence, for a self-conjugate particle all operator coefficients have to be multiplied by a factor of two.}
\begin{lstcpp}
#include "DDCalc.hpp"
int WIMP;
WIMP = DDCalc::InitWIMP();
DDCalc::SetWIMP_NREffectiveTheory(WIMP,50,0.5);
DDCalc::SetNRCoefficient(WIMP,3,0,0.1);
DDCalc::SetNRCoefficient(WIMP,3,1,0.2);
\end{lstcpp}
The second argument of the final function corresponds to the index of the operator to be set, with $q^2 \mathcal{O}_1$ and $q^2 \mathcal{O}_4$ being denoted by $-1$ and $-4$, respectively.

\ddcalc then automatically performs the matching onto the appropriate nuclear response functions, which are evaluated based on the parametrisation and the tabulated values provided in Ref.~\cite{Anand:2013yka}. These tables are provided in the subfolder \path{DDCalc/data/Wbar/} for a range of relevant isotopes. Additional files can be provided to implement additional isotopes, and existing files can be replaced to study form factor uncertainties.

Of course, it is still possible to specify the WIMP coupling structure in the traditional way, e.g.\ by providing the effective couplings for spin-independent (SI) and spin-dependent (SD) interactions with protons and neutrons. In this case, \ddc will by default use the conventional form factors (i.e.\ the Helm form factor for SI interactions and the form factors from Ref.~\cite{Klos} for SD interactions, which can be found in \path{DDCalc/data/SDFF/}). In order to use the form factors from Ref.~\cite{Anand:2013yka} also for standard interactions, one can set the global option \textsf{PreferNewFF} contained in \path{DDCalc/src/DDConstants.f} to \textsf{true}.

Let us finally emphasize that for general non-relativistic operators, the differential event rate depends not only on the conventional velocity integral $\int f(\mathbf{v})/v \, d^3v$ but also on the second velocity integral $\int v \, f(\mathbf{v}) \, d^3v$. As before, these velocity integrals are by default evaluated using the Standard Halo Model (SHM) with parameters that can be set externally. It is however also possible to provide tabulated velocity integrals in order to study velocity distributions that differ from the SHM. An illustration of this feature is provided in the example files \path{DDCalc/examples/DDCalc_exclusionC.cpp} and \path{DDCalc/examples/DDCalc_exclusionF.f90}, which demonstrates how to calculate an exclusion limit for a given WIMP model and a given velocity distribution.

\subsection{Extended detector interface}

The need to implement increasingly complex direct detection experiments has led to substantial extensions of how experiments can be defined in \ddc. The details of this new interface are described in \path{DDCalc/src/DDDetectors.f}, but we review the most important new features here.

First of all, it is now possible to define a number of different signal regions for each experiment and to specify the number of observed events and expected background events for each signal region. The simplest application is the implementation of a binned analysis, but it is also possible to define more complex signal regions, provided they can be characterised by a simple acceptance function $\epsilon(E_R)$, which quantifies the probability that a nuclear recoil with physical recoil energy $E_R$ will lead to a signal within the signal region. \ddc will then determine the expected signal in each signal region and calculate the binned Poisson likelihood. If the expected background in a signal region is set to zero, \ddc will interpret this to mean that the background level is unknown. In this case, the bin will only contribute to the total likelihood if the predicted number of signal events exceeds the number of observed events.  The example files \path{DDCalc/examples/DDCalc_exampleC.cpp} and \path{DDCalc/examples/DDCalc_exampleF.f90} illustrate how the predicted number of events in each signal region, as well as the resulting likelihoods, can be evaluated for specific parameter points.

Alternatively, one can also analyse experiments with unknown backgrounds using the optimum interval method by specifying the bins in such a way that their boundaries correspond to the energies of the observed events. Note that this means that it is typically not possible to use the binned Poisson method and the optimum interval method for the same choice of binning. A user wishing to compare these two analysis strategies should therefore implement them as separate experiments.

A related new feature is that it is now possible in \ddc to specify separate efficiency functions for each element (or indeed each isotope) in the target material. This is necessary for example if the efficiency of analysis cuts depends on the type of recoiling nucleus (as in CRESST) or if the low-energy threshold differs for different elements (as in PICO). For experiments with several different elements and several different signal regions, the number of efficiency functions that need to be specified can potentially be quite large. The preferred way to specify efficiency functions in \ddc is to provide external files with tabulated values, which by default are stored in \path{DDCalc/data/}. An illustration of this new structure can be found in the definition of the CRESST-II experiment (see below).

It is important to emphasize that the grid used to define the efficiency functions is also used to evaluate the other contributions to the differential event rate (i.e., form factors and velocity integrals). The number of grid points used in the definition of the efficiency functions directly influences the computation time and the precision of the result. In particular, it is essential to also provide a sufficiently large number of grid points in energy ranges where the efficiency is approximately constant. The function \textsf{RetabulateEfficiency} in \path{DDCalc/src/DDDetectors.f} can be used to generate a fine efficiency grid from a coarse one, using linear interpolation between the provided values.

\subsection{New experiments}

\ddc ships with a broad range of new experimental analyses. In particular, there are now a number of low-threshold experiments, so that \ddc can now also be used to reliably calculate constraints on light DM. Moreover, we have implemented a number of planned experiments, which can be used to derive projected sensitivities.

\textbf{CRESST-II:}~The CRESST-II results~\cite{Angloher:2015ewa} are based on $52.2 \, \mathrm{kg\,days}$ using the Lise detector module. Our implementation follows Refs.~\cite{Angloher:2017zkf,Kahlhoefer:2017ddj}, i.e., we assume an energy resolution of $\sigma_E = 62\,$eV and take the cut survival probabilities from Ref.~\cite{Angloher:2017zkf}. To avoid unnecessarily fine binning in energy ranges where the expected signal rates are small, we divide the energy range between 0.3 and 5.0 keV into 10 bins of increasing size. In the absence of a background model, we treat all observed events as potential signal events.

\textbf{CDMSlite:}~The analysis of CDMSLite is based on an exposure of $70.14 \, \mathrm{kg\,days}$~\cite{Agnese:2015nto,Agnese:2017jvy}. The energy-dependent signal efficiency is taken from Ref.~\cite{Agnese:2015nto}, which also describes the procedure for converting nuclear recoil energies into electron equivalent energies (eVee). We follow the same approach as in Ref.~\cite{Kahlhoefer:2017ddj} to determine the detector resolution, divide the energy range from $60$ to $500 \, \mathrm{eVee}$ into 10 bins of increasing size and assume no background model.

\textbf{DarkSide-50:}~We implement the results from a search for heavy DM particles in the DarkSide-50 detector based on a total exposure of $19.6 \cdot 10^3 \, \mathrm{kg \, days}$~\cite{Agnes:2018fwg}, taking the energy-dependent acceptance function from Ref.~\cite{Agnes:2018fwg}.

\textbf{PandaX-II:}~Since the most recent data taking period of the PandaX-II experiment (Run 10) has substantially lower background levels than previously analysed data sets~\cite{Tan:2016zwf,Cui:2017nnn}, we implement it as an independent experiment (called \textsf{PandaX\_2017}) rather than simply combining all runs. We use the same detector efficiency for the new data set as for our previous implementation of PandaX-II (see Ref.~\cite{DarkBit}) and assume a background expectation of 1.55 events.\footnote{The expected number of background events is quoted as $1.8\pm0.5$. Assuming the uncertainty in this estimate to be Gaussian, the likelihood is maximized for a background expectation of $1.8 - 0.5^2 = 1.55$ events.} It is then straight-forward to perform a combination of the different data sets by multiplying the individual likelihood functions.

\textbf{XENON1T:}~We use the same implementation of XENON1T~\cite{Aprile:2018dbl} as described in detail in Ref.~\cite{SSDM2}. To reduce background levels, we focus on the central detector region with a mass of 0.65\,t, and consider only events between the median of the nuclear recoil band and the lower $2\sigma$ quantile. We furthermore divide this signal region into two energy bins, which correspond to $\text{S1} \in [3\,\text{PE},\,35\,\text{PE}]$ and  $\text{S1} \in [35\,\text{PE},\,70\,\text{PE}]$. We estimate the expected backgrounds in the two bins to be 0.46 and 0.34 events, respectively, compared to 0 and 2 observed events.

\textbf{LZ:}~Our implementation of the LZ experiment~\cite{Akerib:2015cja} follows Ref.~\cite{Bertone:2017adx}. In particular, we assume an exposure of $5.6 \cdot 10^6 \, \mathrm{kg \, days}$ with a resolution of $\sigma_E / E_R = 0.065 + 0.24 \, (1 \, \mathrm{keV} / E_R)^{1/2}$ and an acceptance of $50\%$ for nuclear recoils. We consider 6 evenly-spaced bins in the range from 6 to 30 keV and assume a background of 0.394 events per bin.

\textbf{PICO-500:}~Our implementation of PICO-500 follows the information provided in Ref.~\cite{PICO500}. PICO-500 plans to employ a C$_3$F$_8$ target with $250 \, \mathrm{L}$ fiducial volume. Six live-months of data will be taken with a low threshold of $3.2 \, \mathrm{keV}$, which we implement using the same acceptance function as for PICO-2L~\cite{Amole:2015lsj}, while 12 live-months will be taken with a threshold of $10 \, \mathrm{keV}$. We treat the two thresholds as two separate bins, in which case the expected backgrounds are 3 and 0.85 events, respectively.

\textbf{DARWIN:}~The DARWIN experiment aims for a total exposure of $7.3 \cdot 10^7 \, \mathrm{kg \, days}$ with 30\% acceptance for nuclear recoils and 99.98\% rejection of electron recoils~\cite{Aalbers:2016jon}. We assume an energy resolution of $\sigma_E / E_R = 0.05 + (0.05 \, \mathrm{keV} / E_R)^{1/2}$~\cite{Schumann:2015cpa} and consider 5 equally-spaced bins between 5 and $20 \, \mathrm{keV}$. The dominant background is due to coherent neutrino-nucleus scattering, which we estimate from Fig.~3 in Ref.~\cite{Schumann:2015cpa}.

\textbf{DarkSide-20k:}~We assume a total exposure of $3.65 \cdot 10^7 \, \mathrm{kg \, days}$ and estimate the energy resolution to be $\sigma_E / E_R = 0.05 + (2 \, \mathrm{keV} / E_R)$~\cite{Aalseth:2017fik}. To model the detector threshold, we implement the acceptance function for the $f_{200}$-cut from Fig.~92 in Ref.~\cite{Aalseth:2017fik}. We divide the energy range between 30 and 80 keV into 10 equally-spaced bins, and assume a background of 0.04 events per bin from instrumental background, as well as a total of 1.6 events (with non-trivial energy dependence) from coherent neutrino scattering, which we obtain by rescaling the results from Ref.~\cite{Billard:2013qya}.

Note that the number of observed events in each bin must be an integer in \ddcalc, so it is typically not possible to set the observed number of events equal to the expected number of events in order to calculate the expected sensitivity of a future experiment. By default, the observed number of events is set to the integer closest to the background expectation, but this introduces a bias for example if there is a large number of bins with less than 0.5 expected background events. To accurately calculate expected sensitivities, one should simulate Poisson fluctuations in each bin, calculate the corresponding exclusion limits, and then construct the median exclusion. For an alternative approach, using Fisher information, we refer to Ref.~\cite{Edwards:2017mnf}.

\begin{figure}[tbp]
 \centering
 \hspace*{-0.6cm}
 \includegraphics[height=0.72\columnwidth]{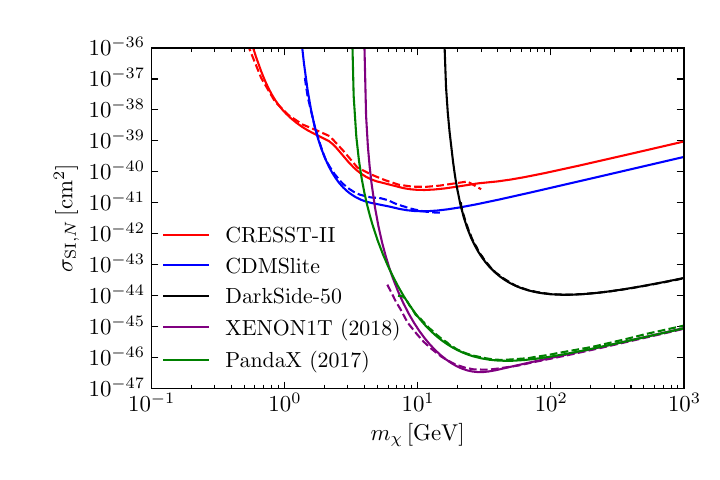}
 \caption{$90\%\,$C.L.\,upper limits on the spin-independent DM-nucleon scattering cross-section from CRESST-II, CDMSlite, DarkSide-50, PandaX-II and XENON1T. The solid curves show the limits obtained using \ddcalc, while the dashed curves correspond to the limits derived by the collaborations~\cite{Angloher:2015ewa,Agnese:2015nto,Agnes:2018fwg,Cui:2017nnn,Aprile:2018dbl}. Note that close to threshold the exclusion limits depend sensitively on the detector response and an accurate modeling in \ddcalc is very challenging.}
 \label{fig::DDCalc_Validation}
\end{figure}

Lastly, in Fig.~\ref{fig::DDCalc_Validation} we show a comparison of the upper bounds on the spin-independent scattering cross-section determined using \ddcalc with the official limits obtained by the respective collaborations. In all cases we find good agreement, validating our implementions of the experimental likelihoods in \ddcalc. Also for the planned experiments described earlier we have confirmed that our sensitivity estimates are in sufficient agreement with the expectations published by the collaborations~\cite{Akerib:2015cja,PICO500,Aalbers:2016jon,Aalseth:2017fik}.

\onecolumn

%%%%%%%%%%%%%%%%%%%%%%%%%%%%%%%%%%%%%%%%%%%%%%%%%%%%%%%%%%%%%%%%%%%%%%%%%%%%%%%%%%%%%%%%%%%%%%%%%%%%%%%%%%%%%%%%%%%%%%%%
\section{Annihilation cross-sections}\label{sec:cross_sections}
In our study, the final states from the DM annihilation include $W^+W^-$, $ZZ$, $\tau^+\tau^-$, $t\bar{t}$, $b\bar{b}$, $c\bar{c}$ and $hh$.
For all final states except $hh$, the DM annihilation proceeds solely via an $s$-channel Higgs exchange. For massive gauge bosons, the annihilation cross-section is
\begin{align}
  \sigma v_\textrm{rel}^\textrm{cms} &= P(X)\frac{s}{8\pi}\delta_i v_i \lambda_{hX}^2 |D_h(s)|^2 \left(1-4x_i+12x_i^2\right) \, ,
\end{align}
where $P(X)$ is defined in Eq.~\eqref{eq:prefactor}, $i=\{W,\,Z\}$,
$\lambda_{hX} \in \{\lambda_{hV}, \, \lambda_{h\chi}/\Lambda_{\chi}, \lambda_{h\psi}/\Lambda_{\psi}\}$,
$\delta_W = 1$, $\delta_Z = 1/2$, $x_i \equiv m_i^2/s$, $v_i=\sqrt{1-4x_i}$, and $|D_h(s)|^2$ is the full squared Higgs propagator given by
\begin{align}
  |D_h (s)|^2 = \frac{1}{\left(s-m_h^2\right)^2 + m_h \Gamma_h (\sqrt{s})} \, .
\end{align}
For fermion final states, the annihilation cross-section is given by
\begin{align}
 \sigma v_\textrm{rel}^\textrm{cms} = P(X)\frac{m_f^2}{4\pi}C_f v_f^3\lambda_{hX}^2|D_h(s)|^2 \, ,
\end{align}
where $C_f$ is a colour factor. For leptons, $C_f = 1$, whereas for quarks, it includes an important 1-loop vertex correction given by \cite{Drees:1990dq}
\begin{equation}
 C_f = 3\left\{1+\left[\frac{3}{2}\log\left(\frac{m_f^2}{s}\right) + \frac{9}{4}\right]\frac{4\alpha_s}{3\pi}\right\} \, .
\end{equation}
For the $hh$ final state, additional contributions appear from the 4-point contact interaction as well as DM exchange in $t$- and $u$-channels.
The annihilation cross-section for $VV\rightarrow hh$ is
\begin{align}
  \sigma v_\textrm{rel}^\textrm{cms} (VV \rightarrow hh) &= \frac{\lambda_{hV}^2v_h}{2304\pi s x_V^4}|D_h(s)|^2 \biggl[ \frac{8\beta v_0^2\lambda_{hV}}{1-2x_h^2}\coth^{-1}\beta \times \nonumber \\
   &\hspace{4mm} \biggl\{ 2s\left(2x_h-1\right)x_V \left(\left(x_h-1\right)\left(2x_h+1\right)-x_\Gamma^2 \right)\left(x_h^2 + 24x_V^3 + 2\left(x_h-1\right)^2 -4\left(2x_h+1\right)x_V^2\right) \nonumber \\
   &\hspace{4mm} - v_0^2\lambda_{hV} \Bigl[ \bigl(3x_h^4 - 8x_h^3x_V - x_h(x_h-4x_V)(8x_V^2+1) - 2x_V(24x_V^3-2x_V+1))(x_h-1)^2 + x_\Gamma^2) \bigr) \Bigr] \biggr\} \nonumber \\
   &\hspace{4mm} + 4s^2x_V^2\left(4x_V\left(3x_V-1\right)+1\right)\left(\left(2x_h+1\right)^2+x_\Gamma^2\right) \nonumber \\
   &\hspace{4mm} -4sx_V\lambda_{hV}v_0^2\left(2x_h\left(2x_V+1\right)+1-6x_V\right)\left(x_h\left(2x_h-1\right)-1-x_\Gamma^2\right) + \frac{\lambda_{hV}^2v_0^4\left(\left(x_h-1\right)^2 + x_\Gamma^2\right)}{x_h^2-4x_Vx_h+x_V} \times \nonumber \\
   &\hspace{4mm} \left(6x_h^4+4x_h^3\left(1-8x_V\right)+x_h^2\left(12x_V\left(4x_V-1\right)+1\right) -64x_V^3x_h + 96x_V^4 + x_V\right)\biggr] \, ,
\end{align}
where the dimensionless quantities $\beta = (1-2 x_h)/(v_h v_V)$ and $x_\Gamma = \Gamma_hm_h/s$, and $v_h$ and $v_V$ are the lab-frame velocities of the Higgs and vector DM, respectively.

Similarly, the annihilation cross-section for $\overline{\chi}\chi\rightarrow hh$ (and equivalently for $\chi \leftrightarrow \psi$) is given by
\begin{align}
  \sigma v_\textrm{rel}^\textrm{cms} (\overline{\chi}\chi \rightarrow hh) &= \left(\frac{\lambda_{h\chi}}{\Lambda_\chi}\right)^2\frac{v_h}{32\pi s} \biggl[
  \left(s - 4\cos^2\xi s x_\chi - 8\cos\xi v_0^2\frac{\lambda_{h\chi}}{\Lambda_\chi}m_\chi\right) + \frac{4\beta s^2 |D_h(s)|^2 v_0^2 \coth^{-1}\beta}{\left(1-2x_h\right)^{2}}\frac{\lambda_{h\chi}}{\Lambda_\chi} \times  \nonumber \\
  &\hspace{4mm} \biggl\{ 2m_\chi\cos\xi\left(2x_h-1\right)\left(x_h\left(2x_h-1\right)-x_\Gamma^2-1\right)\left(8\cos^2\xi x_\chi-2x_h-1\right) \nonumber \\
  &\hspace{4mm} + v_0^2\frac{\lambda_{h\chi}}{\Lambda_\chi}\left(1-4x_h + 6x_h^2 - 16x_\chi \cos^2\xi\left(x_h-1\right)-32\cos^4\xi x_\chi^2\right) \left(\left(x_h-1\right)^2+x_\Gamma^2\right) \biggr\} \nonumber \\
  &\hspace{4mm} + 3s^2|D_h(s)|^2 x_h \left(8\cos\xi v_0^2 \left(x_h-1\right)\frac{\lambda_{h\chi}}{\Lambda_\chi}m_\chi - s\left(x_h+2\right)\left(4\cos^2\xi x_\chi -1\right)\right) \nonumber \\
  &\hspace{4mm} - \left(\frac{\lambda_{h\chi}}{\Lambda_\chi}\right)^2\frac{2v_0^4\left(2x_\chi\left(8\cos^4\xi x_\chi + 1 \right)-8\left(1+\cos^2\xi\right) x_h x_\chi + 3x_h^2\right)}{x_h^2 + x_\chi -4 x_h x_\chi} \nonumber \biggr] \, ,
\end{align}
where $\beta = (1-2x_h)/(v_h v_\chi)$, with $v_\chi$ the lab-frame $\chi$ velocity.

\end{document}